\documentclass[aps,pre,amsmath,amssymb]{revtex4-2}

\usepackage{polski}
\usepackage[english]{babel}
\usepackage{bm}
\usepackage{graphicx}
\usepackage{amsmath}
\usepackage{mathrsfs}

\begin{document}

\newcommand{\uu}[1]{\underline{#1}}
\newcommand{\pp}[1]{\phantom{#1}}
\newcommand{\be}{\begin{eqnarray}}
\newcommand{\ee}{\end{eqnarray}}
\newcommand{\ve}{\varepsilon}
\newcommand{\vp}{\varphi}
\newcommand{\vs}{\varsigma}
\newcommand{\Tr}{{\,\rm Tr\,}}
\newcommand{\Trr}{{\,\rm Tr}}
\newcommand{\pol}{\frac{1}{2}}
\newcommand{\sgn}{{\rm sgn}}
\newcommand{\Mo}{\mho}
\newcommand{\Om}{\Omega}

\title{
Cosmic-time quantum mechanics and the passage-of-time problem}
\author{Marek Czachor}
\affiliation{
Instytut Fizyki i Informatyki Stosowanej,
Politechnika Gdańska, 80-233 Gdańsk, Poland
}

\begin{abstract}
A new dynamical paradigm merging quantum dynamics with cosmology is discussed.
We distinguish between a universe and its background space-time. The universe is here the subset of  space-time defined by $\Psi_\tau(x)\neq 0$, where $\Psi_\tau(x)$ is a solution of a Schr\"odinger equation, $x$ is a point in $n$-dimensional Minkowski space, and  $\tau\geq 0$ is a dimensionless `cosmic time' evolution parameter. We derive the form of the Schr\"odinger equation and show that an empty universe is described by a $\Psi_\tau(x)$ that propagates  towards the future inside of some future-cone $V_+$. The resulting dynamical semigroup is unitary, i.e. $\int_{V_+} d^4x |\Psi_\tau(x)|^2=1$ for $\tau\ge 0$. The initial condition $\Psi_0(x)$ is not localized at $x=0$. Rather, it satisfies the boundary condition $\Psi_0(x)=0$ for $x\not\in V_+$. For $n=1+3$ the support of $\Psi_\tau(x)$ is bounded from the past by the `gap hyperboloid' $\ell^2\sqrt{\tau}=c^2t^2-\bm x^2$, where $\ell$ is a fundamental length. In consequence, the points located between the hyperboloid and the light cone $c^2t^2-\bm x^2=0$ satisfy $\Psi_\tau(x)= 0$, and thus do not belong to the universe. As $\tau$ grows, the gap between the support of $\Psi_\tau(x)$ and the light cone increases. The past thus literally disappears. Unitarity of the dynamical semigroup implies that the universe gets localized in a finite-thickness future-neighborhood of $\ell^2\sqrt{\tau}=c^2t^2-\bm x^2$, simultaneously spreading along the hyperboloid. Effectively, for large $\tau$ the subset occupied by the universe resembles a part of the gap hyperbolid itself, but its thickness $\Delta_\tau$ is nonzero for finite $\tau$. Finite $\Delta_\tau$ implies that the 3-dimensional volume of the universe is finite as well. An approximate radius of the universe, $r_\tau$, grows with $\tau$  due to $\Delta_\tau r_\tau^3=\Delta_0 r_0^3$ and $\Delta_\tau\to 0$. The propagation of $\Psi_\tau(x)$ through space-time matches an intuitive picture of the passage of time. What we regard as the Minkowski-space classical time can be identified with $ct_\tau=\int d^4x\,x^0 |\Psi_\tau(x)|^2$, so $t_\tau$ grows with $\tau$ in consequence of the Ehrenfest theorem, and its present uncertainty can be identified with the Planck time. 
Assuming that at present values of $\tau$ (corresponding to 13-14 billion years) $\Delta_\tau$ and $r_\tau$ are  of the order of the Planck length and the Hubble radius, we estimate that the analogous thickness $\Delta_0$ of the support of $\Psi_0(x)$ is of the order of 1 AU, and $r_0^3\sim (ct_H)^3 \times 10^{-44}$. The estimates imply that the initial volume of the universe was finite and its uncertainty in time was several minutes. Next, we generalize the formalism in a way that incorporates interactions with matter. We are guided by the correspondence principle with quantum mechanics, which should be asymptotically reconstructed for the present values of $\tau$. We argue that Hamiltonians corresponding to the present values of $\tau$  approximately describe quantum mechanics in a conformally Minkowskian space-time. The conformal factor is directly related to $|\Psi_\tau(x)|^2$. 
As a by-product of the construction, we arrive at a new formulation of conformal invariance of $m\neq 0$ fields.
\end{abstract}
\maketitle

\section{Passage of time as a physical problem}
\label{Sec1}

We are taught very early in our education that dynamics in space is equivalent to statics in space-time. As children, we generally have no difficulty with the idea that  a 1-dimensional  motion can be represented by a motionless graph $(t,x_t)$. The  paradigm is easily explainable by the metaphor of a filmstrip, where each moment of time $t$ corresponds to a still frame $x_t$. In a sense, dynamics is not needed in physics.

On the other hand, it would be difficult to find a physical phenomenon whose nature would be experienced by us as directly, as suggestively, and often as dramatically as the passage of time.

The formalism of  invariant-time quantum mechanics partly addresses the issue  \cite{S1,S2,HP,Fanchi1979,HAE,Fanchi1993,Pavsic1991a,Pavsic1991b,Horwitz2015}. Here, one begins with the family of wavefunctions,
$\Psi_\tau(x)$, defined on (1+3)-dimensional Minkowski space (or its generalizations \cite{HP2,Lund}), and satisfying a Schr\"odinger-type equation 
\be
i\dot\Psi_\tau=\mathscr{H}\Psi_\tau, \quad \Psi_\tau=U_{\tau-\tau_0}\Psi_{\tau_0}.\label{1}
\ee
The normalization is $\int d^4x |\Psi_\tau(x)|^2=1$. The resulting dynamics is no longer an equivalent of statics in four dimensions. But does it really match our intuition of passage of time, where the past is disappearing and the future has not yet happened? 

So, consider the following sequence of syllogisms:

An event cannot happen if its probability is zero. Probability of $x$ is zero if $|\Psi_\tau(x)|^2=0$. An event that could happen at $\tau_1$ disappears  at $\tau_2$ if 
$\Psi_{\tau_1}(x)\neq 0$ evolves into $\Psi_{\tau_2}(x)=0$. $\Psi_\tau(x)$ describes a passage of time if its support is restricted from the past by a spacelike hypersurface propagating toward the future. 

The above postulates should be supplemented by the asymptotic one: For times of the order of 13-14 billion years since the origin of the cosmic evolution the support of $\Psi_\tau(x)$ should be `practically' indistinguishable from a spacelike hyperplane, at least locally (say, at galaxy scale).

We will therefore define {\it a\/} universe as a collection of those events $x$ in Minkowski space $\cal M$ that satisfy $\Psi_{\tau}(x)\neq 0$ for a certain solution of (\ref{1}), for some $\mathscr{H}$. We will determine $\mathscr{H}$ by the condition that for very large $\tau$ the probability density $|\Psi_\tau(x)|^2$ will be concentrated in a neighborhood of a hyperbolic subspace of $\cal M$. This subspace will propagate in $\cal M$ toward the future. For smaller $\tau$, instead of a spacelike hyperboloid, what we find is a finite-thickness $n$-dimensional quantum membrane propagating through the Minkowski space of the same dimension. The membrane simultaneously spreads along spacelike directions and shrinks along the timelike ones. The two processes balance each other, making the dynamics unitary.  Asymptotically, for large cosmic times, the dynamics becomes similar to Dirac's point form \cite{Dirac1949}.

Notice that we speak here of a {\it neighborhood\/} of the hyperbolic subspace, and not just the hyperbolic subspace itself. What it means is that the asymptotic (empty) universe is an $n$-dimensional subset of the $n$-dimensional $\cal M$, and not its $(n-1)$-dimensional submanifold. Our membrane resembles a true material membrane of finite thickness, and not just its idealized  $(n-1)$-dimensional mathematical representation.

The choice of hyperbolic geometry is motivated by reasons of symmetry, isotropy, unboundedness, and homogeneity of the asymptotic universe.  Regarded as $(n-1)$-dimensional manifolds, hyperbolic spaces are isotropic homogeneous spaces of constant negative curvature \cite{Helgason}. For $n=4$ they are examples of spatial sections of a Robertson-Walker space-time  \cite{Weinberg}. Alternatively, they are spatial sections of a Milne universe \cite{Milne,Milne2,Bondi,Chodorowski,Kutschera,Vishwakarma2013,NGS,RM,VishwakarmaNarlikar,Vishwakarma2020,Zaninetti}. Hyperbolic spaces are natural candidates for universes that are either completely empty, or filled with test matter (identified by Milne with galaxies). In particular, a universe filled with several interacting atoms, say, could be described by a hyperbolic space. 

The classical Kepler problem was solved in 3-dimensional hyperbolic space in \cite{Chernikov}.
Kepler's problem is apparently also the first quantum problem solved in hyperbolic space \cite{Infeld,Kozlov}. Quantum mechanical harmonic oscillator on various spaces of constant curvature  is another example \cite{Carinena}. Eigenfunction expansions on hyperboloids and cones of various metric signatures can be found in \cite{Raczka}, whereas the special case of $n=1+3$ appeared in \cite{Dirac1945}, and in more complete forms in \cite{Naimark} and \cite{Zmuidzinas}. A more recent study can be found in \cite{1,2}.

It is known that the Milne model fits observational data for Type Ia supernovae just as well as the $\Lambda$CDM model \cite{DE1,DE2}, at least when one considers the Hubble diagram for distance modulus vs. redshift \cite{NGS,RM}. The differences between $\Lambda$CDM and Milne's models become visible if one switches to `model-independent' scale factor vs. cosmological time plots \cite{RM,RM2014}, but one should bear in mind that the notion of  `model-independence' is referred here to a specific class of models which do not include the formalism we discuss in the present paper. Therefore, we withhold for the time being a final opinion on the possible agreement or disagreement of our model with the observational data.

A more technical and detailed outline of the construction is given in the next Section. An example of $n=1+1$  illustrates our main intuitions. Sec.~\ref{Sec3} is central to the paper. The construction of $U_\tau=e^{-i \tau{\cal H}_0 }$ is given there step by step. Sec.~\ref{Sec4} plays a role of a cross-check of the construction from Sec.~\ref{Sec3}. Sec.~\ref{Sec5} is devoted to spectral properties of ${\cal H}_0$. Sections~\ref{Sec6}--\ref{Sec8} deal with various properties of the universe which we identify with the support of $\Psi_\tau(x)$, a solution of the Schr\"odinger equation. 

A very preliminary analysis of such a dynamics for $n=1+1$ can be found in \cite{MCAP}. A disadvantage of the approach from \cite{MCAP} was that it crucially depended on properties of $(1+1)$-dimensional Minkowski space, treated as a toy model. The new formalism is independent of the dimension of the background space.

In Sec.~\ref{Sec int1} we begin discussion of matter fields and justify the form of the total Schr\"odinger-picture Hamiltonian. In particular, we point out that what we regard as matter-field total Hamiltonian in our present-day universe is essentially an interaction Hamiltonian. In Sec.~\ref{Sec int2} we discuss the link between the {\it averaged-over-reservoir\/} interaction Hamiltonian and the resulting effective geometry of the universe. The geometry depends on the initial condition for 
$\Psi_\tau(x)$ and is encoded in the structure of spinor covariant derivative. We argue in Sec.~\ref{Sec int3} that the most natural choice of the derivative is the one with non-vanishing torsion. We compare our construction with the classic results of Penrose on torsion and complex conformal transformations. As a by product we arrive at a connection that leads to a new perspective on the old problem of conformal invariance of massive fields. These ideas are explicitly checked on the example of the Dirac equation in Sec.~\ref{Sec int4}.

In Section~\ref{Sec 1+1} we conclude the paper by a simple toy-model analysis performed in $1+1$ dimensions. All the essential elements of the construction can be followed once again step by step.

The last Section summarizes our assumptions and intuitions, both physical and mathematical, and outlines possibilities of further generalizations of the formalism.

\section{Outline of the construction}
\label{Sec2}

Consider the Minkowski space $\cal M$ in $n$ dimensions with the metric of signature $(+,-,\dots,-)$. We are basically interested in the physical case $n=1+3$, but $n=1+1$ is often needed for graphical illustrations of the construction. Consider an arbitrary $X^{\mu}\in \cal M$ and its future cone $\bar V_+\subset \cal M$, i.e. $x^{\mu}\in \bar V_+$ if $x^{\mu}-X^{\mu}$ is future-pointing and timelike or null. The interior of $\bar V_+$ is denoted by $V_+$, so $\partial V_+=\bar V_+\setminus V_+$ is  the future light-cone of $X^{\mu}$. 
In what follows, we simplify notation by setting $X^{\mu}=0$, but bear in mind that the origin is in fact arbitrary and subject to a Poincar\'e transformation. So, the Poincar\'e group (as well as its unitary representations) is implicitly present as a symmetry group of the background Minkowski space. 

We will concentrate on the Hilbert space of square-integrable functions ${\cal M}\ni x^{\mu}\mapsto \Psi(x)\in \mathbb{C}$, which are assumed to vanish if $x^0\to \infty$, and if $x^{\mu}\in {\cal M}\setminus V_+$. Notice that the wave functions vanish on the boundary $\partial V_+$, so the arguments of $\Psi(x)$ are effectively future-timelike. The scalar product is
\be
\langle f|g\rangle 
&=&
\int_{\cal M}d^nx \overline{f(x)}g(x)
=
\int_{V_+}d^nx \overline{f(x)}g(x).\label{<f|g>}
\ee
For $x^{\mu}\in \bar V_+$ we denote $\mathtt x^2=x_{\mu}x^{\mu}=(x_0)^2-(x_1)^2-\dots-(x_{n-1})^2$. The boundary condition means that we consider wave functions that vanish for $x_{\mu}x^{\mu}\le 0$, and for  $x_{\mu}x^{\mu}> 0$ but belonging to the past cone  $x_0<0$. 

Our goal is to construct a unitary dynamics $\Psi_\tau(x)=U_{\tau-\tau_0}\Psi_{\tau_0}(x)$, fulfilling the following two requirements:

1) $\mathscr{Z}=\max_x \{|\Psi_\tau(x)|^2\}=\max_x \{|\Psi_{\tau_0}(x)|^2\}$ for any $\tau$, $\tau_0$. The condition means that $\mathscr{Z}$ is the maximal value of $|\Psi_\tau(x)|^2$, which is both relativistically and dynamically invariant. In a wider perspective, such a $\mathscr{Z}$ will play a role of a renormalization constant, while $|\Psi_\tau(x)|^2/\mathscr{Z}$ will be a cutoff function whose support defines the region of space-time that will be identified with the universe itself. So, the universe is a $\tau$-dependent subset of the background Minkowski space. 

2) For $\tau\to \infty$ the support of $|\Psi_\tau(x)|^2$ gets concentrated in a neighborhood of a proper-time hyperboloid 
$a_\tau^2=x_{\mu}x^{\mu}$, for some $a_\tau$, $\lim_{\tau\to\infty}a_\tau=\infty$. We will make the condition mathematically precise later; the basic intuition behind it is that, for large times, the probability density on space-time is concentrated in a neighborhood of a spacelike surface propagating toward the future. The propagating support of $|\Psi_\tau(x)|^2$ behaves as if it scanned $V_+$ by a spacelike effective foliation of a finite but decreasing-in-$\tau$ timelike thickness $\Delta_\tau$, 
$\lim_{\tau\to\infty}\Delta_\tau=0$. The latter, when combined with $\mathscr{Z}=$~const, implies that  $|\Psi_\tau(x)|^2$  spreads along spacelike directions, a property we interpret as expansion of our universe. More precisely, this will be one of the manifestations of the expansion, not necessarily the observable one. In effect, the asymptotic dynamics becomes analogous to Dirac's point-form one \cite{Dirac1949}.

The assumptions will lead to the semigroup \cite{Davies}
\be
\Psi_{\tau}(x)
&=&
\Psi_{\tau_0}\left(
\left(\frac{\mathtt x^n-(a_\tau)^n+(a_{\tau_0})^n}{\mathtt x^n}\right)^{1/n}x
\right)\label{2''}\\
&=&
e^{-i\left((a_{\tau})^n-(a_{\tau_0})^n\right){\cal H}_0/\ell^n}
\Psi_{\tau_0}(x),\label{2}
\ee
for $(a_\tau)^n-(a_{\tau_0})^n<\mathtt x^n$,
and 
\be
\Psi_{\tau}(x)
&=&
0,\label{2'''}
\ee
for $0\le \mathtt x^n\le (a_\tau)^n-(a_{\tau_0})^n$,
\be
{\cal H}_0&=&
-
\frac{\ell^n}{n\mathtt x^n}x^{\mu}\, i\partial_{\mu}
=
-i
\ell^n\frac{\partial}{\partial(\mathtt{x}^n)}
,\label{Omega}
\ee
where $\ell$ is a constant (the Planck length, say). 
Formula (\ref{2}) shows that the parameter that plays a role of a `quantum time' is here given by $(a_{\tau})^n$. 
It is most natural (and simplest) to work with 
\be
(a_{\tau})^n &=&\ell^n\tau,\label{a6}\\
{\cal H}_0 &=& V^\mu P_\mu,\\
V^{\mu}
&=&
\frac{\ell^{n-1}}{n\mathtt x^n}x^{\mu}, \quad \partial_{\mu} V^{\mu}(x)=0,\\
P_\mu &=& -i\ell \partial_\mu,\\
\Psi_{\tau}(x)
&=&
\Psi_{\tau_0}\left(
\left(\frac{\mathtt x^n-\ell^n\tau+\ell^n\tau_0}{\mathtt x^n}\right)^{1/n}x
\right)\label{2'}\\
&=&
e^{-i(\tau-\tau_0){\cal H}_0}
\Psi_{\tau_0}(x),\label{3'}
\ee
The parameter $\tau$ is then dimensionless and non-negative. The Hamiltonian ${\cal H}_0$ is dimensionless as well.

Hamiltonian ${\cal H}_0$ is, up to the denominator, a dilatation operator, which is not that surprising in the context of cosmology \cite{BigBounce}. It is clear that, due to the distinguished role played by dilatations,  the resulting formalism has formal similarities to Klauder's affine quantization \cite{Klauder0,Klauder1,Klauder2}. More importantly, 
${\cal H}_0$ generates translations in $n$th power of $\mathtt{x}$, a fact explaining why the dynamics involves a unitary representation of a semigroup of translations in $\mathbb{R}_+$. 

As opposed to algebraic  quantization paradigms (canonical, affine, etc.) we do not begin with with a classical theory, find its Poisson-bracket Lie algebra, and then look for its representations. Our procedure concentrates on the very process of `flow of time' that we envisage as a propagation of a wave packet of the universe through  background space-time. There is, though, a classical element that 
relates our quantum dynamics to more standard Milne-type cosmology: The support of the propagating wave packet is bounded from below (that is, from the past) by a typical Milnean hyperboloid propagating toward the future. As $\tau$ tends to plus-infinity, the wave function concentrates in a future-neighborhood of the propagating hyperboloid. 

As we can see, our dynamics is not just statics in space-time. We indeed have a flow of time, with the past disappearing in the deepest ontological sense, and the future not yet existing. The notion of `now' is smeared out, but becomes more and more concrete as the cosmic time flows toward the future.

\begin{figure}
\includegraphics[width=8 cm]{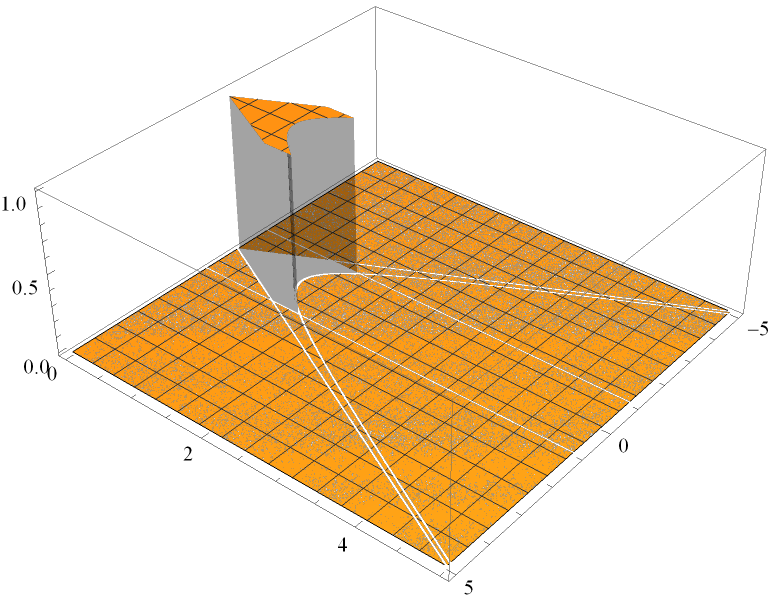}
\includegraphics[width=8 cm]{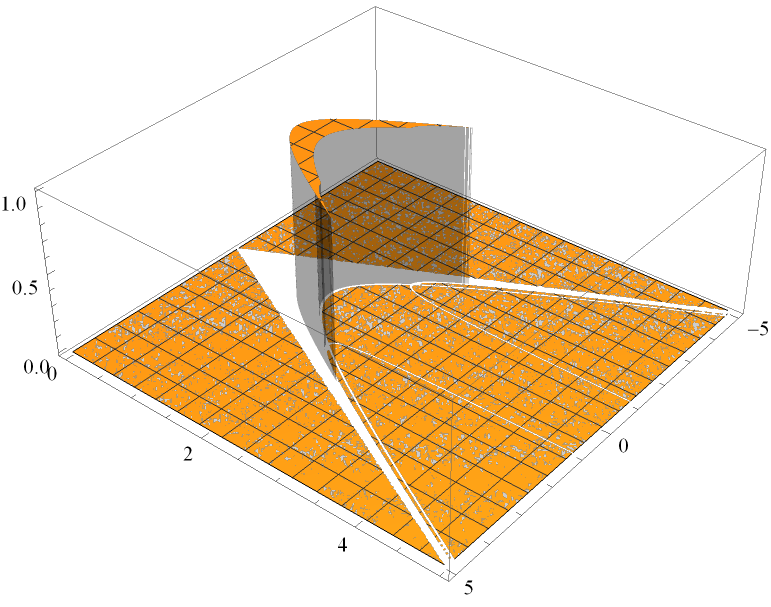}
\caption{Plot of (\ref{2}) with the initial condition (\ref{9}) at $\tau=\tau_0=0$ (left), and its evolved version for $\tau=1$ (right) in a two-dimensional Minkowski space, in units where $\ell=1$. $\Psi_\tau(x)$ at $\tau=1$ is thinner and wider than $\Psi_\tau(x)$ at $\tau=0$. A space-time gap occurs between the support of $\Psi_1(x)$ and the light cone. In $1+3$ dimensions the hyperboloid that determines the gap is given by $\ell^2\sqrt{\tau}=c^2t^2-\bm x^2$, as contrasted with 
$c^2t^2-\bm x^2\sim \tau^2$ one might naively expect.}
\label{Fig1}
\end{figure}

Continuity equation $\partial_{\mu} V^{\mu}(x)=0$ implies that ${\cal H}_0$ is symmetric,
\be
\langle f|{\cal H}_0 g\rangle = \langle {\cal H}_0 f|g\rangle.
\ee
Let us note that the support of $\Psi_\tau$ consists of those $x\in \bar V_+$ that satisfy 
\be
(a_{\tau})^n-(a_{\tau_0})^n\le \mathtt x^n.
\ee
With growing $a_{\tau}$ the support of $\Psi_\tau$ shrinks, creating a space-time gap between the region of non-zero probability density 
$|\Psi_\tau(x)|^2$ and the boundary $\partial V_+$. Fig.~\ref{Fig1} illustrates the effect in $n=1+1$, for (\ref{a6}) and  $\Psi_\tau(x)$ given by (\ref{2''}), (\ref{2'''}), 
with the initial condition
\be
\Psi_0(x)
=
\left\{
\begin{array}{cl}
1 & \textrm{for $|x^1|<1$,  $(x^0)^2-(x^1)^2<1$ , $ x_0>0$}\\
0& \textrm{otherwise} 
\end{array}
\right.
\label{9}
\ee
We tacitly assume that the jumps in (\ref{9}) approximate some smooth function, so that (\ref{2}) is applicable as well.

\section{Justification of the form of $U_\tau$ for the empty universe}
\label{Sec3}

Let $u^{\mu}=x^{\mu}/\mathtt x$, $u^0>0$, be a future-pointing world-velocity. Assume 
$ 
\langle\Psi_\tau|\Psi_\tau\rangle=
\langle\Psi_{\tau_0}|\Psi_{\tau_0}\rangle
$
for any $0\le\tau_0\le \tau$. Explicitly,
\be
\langle\Psi_\tau|\Psi_\tau\rangle
&=&
\int_{V_+} d^nx |\Psi_\tau(x)|^2
=
\int_{0}^\infty d\mathtt x\, {\mathtt x}^{n-1}\rho_\tau(\mathtt x)
\nonumber\\
&=&
\int_{0}^\infty d\mathtt x\, {\mathtt x}^{n-1}\rho_{\tau_0}(\mathtt x).
\ee
We have introduced the probability density
\be
\rho_\tau(\mathtt x) &=& \int_{u^2=1} du |\Psi_\tau(\mathtt x u)|^2,
\ee
where $du$ is a measure on the world-velocity hyperboloid. 

Assuming the support of $\rho_\tau(\mathtt x)$ is contained in $[a_\tau,b_\tau]\subset \mathbb{R}_+$ (as in Fig.~\ref{Fig1}), we arrive at the condition that has to be satisfied by both $\Psi_\tau(x)$ and $\Psi_{\tau_0}(x)$,
\be
\int_{a_\tau}^{b_\tau} d\mathtt x\,\mathtt x^{n-1}\rho_\tau(\mathtt x)
=
\int_{a_{\tau_0}}^{b_{\tau_0}} d\mathtt x\,\mathtt x^{n-1}\rho_{\tau_0}(\mathtt x).
\ee
Changing variables, $t={\mathtt x}^n$, and denoting $\varrho_\tau(t)=\rho_\tau(t^{1/n})$, we find an equivalent form,
\be
\int_{a_\tau^{n}}^{b_\tau^{n}} d t\varrho_\tau(t)
=
\int_{a_{\tau_0}^{n}}^{b_{\tau_0}^{n}} d t\varrho_{\tau_0}(t).\label{24}
\ee
Now, it is enough to find another change of variables, $t\mapsto t_0$, in a way that $a_{\tau}^{n} \leq t \leq b_{\tau}^{n}$ implies 
$a_{\tau_0}^{n} \leq t_0 \leq b_{\tau_0}^{n}$. Assuming the affine relation,
\be
t &=& At_0+B,\\
a_{\tau}^{n} &=& A a_{\tau_0}^{n}+B,\\
b_{\tau}^{n} &=& A b_{\tau_0}^{n}+B,
\ee
we obtain
\be
t &=&
\frac{b_{\tau}^{n}-a_{\tau}^{n}}{b_{\tau_0}^{n}-a_{\tau_0}^{n}}
t_0
+
\frac{-a_{\tau}^{n}b_{\tau_0}^{n}
+
b_{\tau}^{n} a_{\tau_0}^{n}}{a_{\tau_0}^{n}-b_{\tau_0}^{n}}.
\ee
Applying the new variables to the left side of (\ref{24}), 
$dt =
\frac{b_{\tau}^{n}-a_{\tau}^{n}}{b_{\tau_0}^{n}-a_{\tau_0}^{n}}
dt_0$, 
we arrive at
\be
\int_{a_{\tau_0}^{n}}^{b_{\tau_0}^{n}} d t_0 
\frac{b_{\tau}^{n}-a_{\tau}^{n}}{b_{\tau_0}^{n}-a_{\tau_0}^{n}}
\varrho_\tau(At_0+B)
=
\int_{a_{\tau_0}^{n}}^{b_{\tau_0}^{n}} d t_0\varrho_{\tau_0}(t_0)
\ee
and
\be
\frac{b_{\tau}^{n}-a_{\tau}^{n}}{b_{\tau_0}^{n}-a_{\tau_0}^{n}}
\varrho_\tau(t)
=
\varrho_{\tau_0}(t_0)
=
\varrho_{\tau_0}\big(A^{-1}(t-B)\big)
\ee
In order to guarantee the dynamical invariance of $\mathscr{Z}=\max_x \{|\Psi_\tau(x)|^2\}$, we demand
\be
\frac{b_{\tau}^{n}-a_{\tau}^{n}}{b_{\tau_0}^{n}-a_{\tau_0}^{n}}=1.
\ee
The latter implies that
\be
b_{\tau}^{n}-a_{\tau}^{n}=b_{\tau_0}^{n}-a_{\tau_0}^{n}=C
\ee
is a constant, and thus  $b_{\tau}=(a_{\tau}^{n}+C)^{1/n}$. 
Then
\be
\varrho_\tau(t)
=
\varrho_{\tau_0}(t_0)
=
\varrho_{\tau_0}(t-B),\label{n36}
\ee
where
\be
t &=& t_0+B=
t_0
+
a_{\tau}^{n}
-
a_{\tau_0}^{n}.\label{n37}
\ee
It is clear that 
$a_{\tau}^{n} \leq t \leq a_{\tau}^{n}+C$ is equivalent to
$a_{\tau_0}^{n} \leq t_0 \leq a_{\tau_0}^{n}+C$.

As the final step we note that 
\be
\rho_\tau(\mathtt x)
&=&
\varrho_\tau({\mathtt x}^n)
=
\varrho_{\tau_0}({\mathtt x}^n-B)
\nonumber\\
&=&
\varrho_{\tau_0}\big([({\mathtt x}^n-B)^{1/n}]^n\big)
=
\rho_{\tau_0}\big(({\mathtt x}^n-B)^{1/n}\big)
\ee
is equivalent to
\be
&{}&\int_{u^2=1} du |\Psi_\tau(\mathtt x u)|^2
\nonumber\\
&{}&\pp=
=
 \int_{u^2=1} du \left|\Psi_{\tau_0}\left(\big(
{\mathtt x}^n-a_{\tau}^{n}+a_{\tau_0}^{n}
\big)^{1/n}u\right)\right|^2.
\ee
We will now show that 
\be
\Psi_\tau(\mathtt x u)
&=&
\Psi_{\tau_0}\left(\big(
{\mathtt x}^n-a_{\tau}^{n}+a_{\tau_0}^{n}
\big)^{1/n}u\right)
\ee
satisfies all our desiderata.

Firstly, for $x^{\mu}={\mathtt x} u^{\mu}$ we obtain
\be
\Psi_\tau(x)
&=&
\Psi_{\tau_0}\left(\big(
{\mathtt x}^n-a_{\tau}^{n}+a_{\tau_0}^{n}
\big)^{1/n}x/\mathtt x\right)
\\
&=&
\Psi_{\tau_0}\left(
\left(\frac{\mathtt x^n-a_{\tau}^{n}+a_{\tau_0}^{n}}{\mathtt x^n}\right)^{1/n}x
\right)
\ee
which coincides with (\ref{2''}). For $a_{\tau}^{n}=\ell^n \tau$, $\tau \ge 0$, and $\tau_0=0$:
\be
\Psi_{\tau}(x)
&=&
\Psi_{0}\left(
\left(\frac{\mathtt x^n-\ell^n\tau}{\mathtt x^n}\right)^{1/n}x
\right)=\Psi_0\big(x(\tau)\big),\\
x_{\mu}(\tau)
&=&
\left(\frac{\mathtt x^n-\ell^n\tau}{\mathtt x^n}\right)^{1/n}x_{\mu}.
\ee
It is enough if we show that for any analytic function $\Psi_0(x)$ we can write
\be
\Psi_0\big(x(\tau)\big)
&=&
e^{-i\tau {\cal H}_0}\Psi_0(x).
\ee
We begin with the monomial
\be
\Psi_0(x)
=
x_{\mu_1}^{N_1}\dots x_{\mu_k}^{N_k},\quad N_1+\dots +N_k=N.
\ee
It is clear that
\be
x_{\mu_1}(\tau)^{N_1}\dots x_{\mu_k}(\tau)^{N_k}
&=&
e^{-i\tau {\cal H}_0}
x_{\mu_1}^{N_1}\dots x_{\mu_k}^{N_k},
\ee
holds true if and only if
\begin{widetext}
\be
\frac{d^m}{d\tau^m}x_{\mu_1}(\tau)^{N_1}\dots x_{\mu_k}(\tau)^{N_k}\Big|_{\tau=0}
&=&
(-i{\cal H}_0)^m
x_{\mu_1}^{N_1}\dots x_{\mu_k}^{N_k},
\ee
for any $m\in \mathbb{N}$. We begin with
\be
x_{\mu_1}(\tau)^{N_1}\dots x_{\mu_k}(\tau)^{N_k}
&=&
\left(\frac{\mathtt x^n-\ell^n\tau}{\mathtt x^n}\right)^{N/n}
x_{\mu_1}^{N_1}\dots x_{\mu_k}^{N_k}\\
&=&
\frac{\left(\mathtt x^n-\ell^n\tau\right)^{N/n}}{(\mathtt x^2)^{N/2}}
x_{\mu_1}^{N_1}\dots x_{\mu_k}^{N_k},
\ee
\be
\frac{d^m}{d\tau^m}x_{\mu_1}(\tau)^{N_1}\dots x_{\mu_k}(\tau)^{N_k}\Big|_{\tau=0}
&=&
(-\ell^n)^m\frac{N}{n}\left(\frac{N}{n}-1\right)\dots \left(\frac{N}{n}-m+1\right)
\frac{\left(\mathtt x^n-\ell^n\tau\right)^{N/n-m}}{(\mathtt x^2)^{N/2}}
x_{\mu_1}^{N_1}\dots x_{\mu_k}^{N_k}\Big|_{\tau=0}\\
&=&
(-\ell^n)^m\frac{N}{n}\left(\frac{N}{n}-1\right)\dots \left(\frac{N}{n}-m+1\right)
(\mathtt x^2)^{\frac{n}{2}\frac{N}{n}-\frac{n}{2}m-\frac{N}{2}}
x_{\mu_1}^{N_1}\dots x_{\mu_k}^{N_k}\\
&=&
(-\ell^n)^m\frac{N}{n}\left(\frac{N}{n}-1\right)\dots \left(\frac{N}{n}-m+1\right)
(\mathtt x^2)^{-nm/2}
x_{\mu_1}^{N_1}\dots x_{\mu_k}^{N_k}.\label{51}
\ee
On the other hand, by Euler's homogeneity theorem,
\be
(-i{\cal H}_0)^m
x_{\mu_1}^{N_1}\dots x_{\mu_k}^{N_k}
&=&
\left(
-\frac{\ell^n}{n\mathtt x^n}x^{\mu}\partial_{\mu}
\right)^m
x_{\mu_1}^{N_1}\dots x_{\mu_k}^{N_k}\\
&=&
\frac{(-\ell^n)^m}{n^m}\left(
\frac{1}{\mathtt x^n}x^{\mu}\partial_{\mu}
\right)^{m-1}
\frac{N}{\mathtt x^n}
x_{\mu_1}^{N_1}\dots x_{\mu_k}^{N_k}\\
&=&
\frac{(-\ell^n)^m}{n^m}\left(
\frac{1}{\mathtt x^n}x^{\mu}\partial_{\mu}
\right)^{m-2}
\frac{N(N-n)}{(\mathtt x^2)^{2n/2}}
x_{\mu_1}^{N_1}\dots x_{\mu_k}^{N_k}\\
&=&
\frac{(-\ell^n)^m}{n^m}\left(
\frac{1}{\mathtt x^n}x^{\mu}\partial_{\mu}
\right)^{m-3}
\frac{N(N-n)(N-2n)}{(\mathtt x^2)^{3n/2}}
x_{\mu_1}^{N_1}\dots x_{\mu_k}^{N_k}\\
&\vdots&\nonumber\\
&=&
(-\ell^n)^m\frac{N}{n}\left(\frac{N}{n}-1\right)\dots \left(\frac{N}{n}-m+1\right)
(\mathtt x^2)^{-nm/2}
x_{\mu_1}^{N_1}\dots x_{\mu_k}^{N_k},
\ee
which coincides with (\ref{51}).  So, this step is proved. The Maclaurin expansion ends the proof for the monomial,
\be
x_{\mu_1}(\tau)^{N_1}\dots x_{\mu_k}(\tau)^{N_k}
&=&
\sum_{m=0}^\infty
\frac{\tau^m}{m!}\frac{d^m}{d\tau^m}x_{\mu_1}(\tau)^{N_1}\dots x_{\mu_k}(\tau)^{N_k}\Big|_{\tau=0}\\
&=&
\sum_{m=0}^\infty
\frac{\tau^m}{m!}
(-i{\cal H}_0)^m
x_{\mu_1}^{N_1}\dots x_{\mu_k}^{N_k}
=
e^{-i\tau {\cal H}_0}
x_{\mu_1}^{N_1}\dots x_{\mu_k}^{N_k}.
\ee
\end{widetext}
By linearity the proof is extended to any analytic 
\be
\Psi_0(x)=\sum_{N_1\dots N_k}\Psi^{\mu_1\dots \mu_k}_{N_1\dots N_k}x_{\mu_1}^{N_1}\dots x_{\mu_k}^{N_k}.
\ee
For $0\le \tau_0\le \tau$, we can write  
\be
\Psi_\tau(x)
&=&
e^{-i\tau {\cal H}_0}
\Psi_0(x)
=
e^{-i(\tau-\tau_0+\tau_0) {\cal H}_0}
\Psi_0(x)
\nonumber\\
&=&
e^{-i(\tau-\tau_0) {\cal H}_0}
\Psi_{\tau_0}(x)=U_{\tau-\tau_0}\Psi_{\tau_0}(x).
\ee
Being linear and norm-preserving, $U_{\tau-\tau_0}$ is unitary.

\section{Direct proof of unitarity of $U_\tau$}
\label{Sec4}

To remain on a safe side we have assumed $0\le \tau_0\le \tau$, or more generally $0\le a_{\tau_0}\le a_{\tau}$. However, do we really need 
$a_{\tau_0}\le a_{\tau}$? Let us investigate this point in more detail. 

It is instructive to directly verify $\frac{d}{d\tau}\langle \Psi_\tau|\Phi_\tau\rangle=0$ for functions $\Psi_\tau(x)$, $\Phi_\tau(x)$ that vanish outside of $[a_\tau,(a_\tau^n+C)^{1/n}]$. Denote
\be
\rho_\tau(\mathtt x)
=\int_{u^2=1}du \,\overline{\Psi_\tau(\mathtt x u)}\Phi_\tau(\mathtt x u).
\ee
Then
\be
&{}&\frac{d}{d\tau}\int d^4x \overline{\Psi_\tau(x)}\Phi_\tau(x)
\nonumber\\
&{}&
\pp{\frac{d}{d\tau}}
=
\frac{d}{d\tau}
\int_{a_{\tau}}^{(a_\tau^n+C)^{1/n}} d\mathtt x \,\mathtt x^{n-1}
\rho_{\tau_0}\left(\big(
{\mathtt x}^n-a_{\tau}^{n}+a_{\tau_0}^{n}
\big)^{1/n}\right)
\nonumber\\
&{}&
\pp{\frac{d}{d\tau}}
=
\frac{1}{n}\frac{d(a_\tau^n)}{d\tau}\frac{d}{d(a_\tau^n)}
\int_{a_{\tau}^n}^{a_\tau^n+C} dt \,
\rho_{\tau_0}\left(\big(
t-a_{\tau}^{n}+a_{\tau_0}^{n}
\big)^{1/n}\right).
\nonumber
\ee
Employing
\be
&{}&
\frac{d}{dx}
\int_x^{x+c}dy\, f(x,y)
\nonumber\\
&{}&
\pp{\frac{d}{d\tau}}
=
\int_x^{x+c}dy \frac{\partial f(x,y)}{\partial x}
+
f(x,x+c)-f(x,x)\nonumber
\ee
we find
\begin{widetext}
\be
\frac{d}{d\tau}\langle \Psi_\tau|\Phi_\tau\rangle
&=&
\frac{1}{n}\frac{d(a_\tau^n)}{d\tau}
\left[
-\int_{a_{\tau}^n}^{a_\tau^n+C} dt \,
\frac{d}{dt}
\rho_{\tau_0}\left(\big(
t-a_{\tau}^{n}+a_{\tau_0}^{n}
\big)^{1/n}\right)
+
\rho_{\tau_0}\left(\big(
C+a_{\tau_0}^{n}
\big)^{1/n}\right)
-
\rho_{\tau_0}(|a_{\tau_0}|)
\right]
=0.
\ee
\end{widetext}
Both $C+a_{\tau}^{n}$ and $a_{\tau}^{n}$ have to be non-negative for any $\tau$, but $(a_\tau)^n-(a_{\tau_0})^n$ can be of either sign. 

One can prove $\frac{d}{d\tau}\langle \Psi_\tau|\Phi_\tau\rangle=0$ also under a slightly different condition. Namely, assume (\ref{2'''}), 
\be
\Psi_{\tau}(x)
&=&
0,
\ee
for $0\le \mathtt x^n\le (a_\tau)^n-(a_{\tau_0})^n$. Then
\begin{widetext}
\be
\frac{d}{d\tau}\langle \Psi_\tau|\Phi_\tau\rangle
&=&
\frac{d}{d\tau}
\int_{\left(a_{\tau}^{n}-a_{\tau_0}^{n}\right)^{1/n}}^\infty d\mathtt x \,\mathtt x^{n-1}
\rho_{\tau_0}\left(\big(
{\mathtt x}^n-a_{\tau}^{n}+a_{\tau_0}^{n}
\big)^{1/n}\right)
\\
&=&
\frac{1}{n}
\frac{d(a_{\tau}^{n}-a_{\tau_0}^{n})}{d\tau}
\frac{d}{d(a_{\tau}^{n}-a_{\tau_0}^{n})}
\int_{a_{\tau}^{n}-a_{\tau_0}^{n}}^\infty dt \,
\rho_{\tau_0}\left(\big(
t-(a_{\tau}^{n}-a_{\tau_0}^{n})
\big)^{1/n}\right)
\\
&=&
\frac{1}{n}
\frac{d(a_{\tau}^{n}-a_{\tau_0}^{n})}{d\tau}
\left[
-
\rho_{\tau_0}(0)
-
\int_{a_{\tau}^{n}-a_{\tau_0}^{n}}^\infty dt \,
\frac{d}{dt}
\rho_{\tau_0}\left(\big(
t-(a_{\tau}^{n}-a_{\tau_0}^{n})
\big)^{1/n}\right)
\right]\\
&=&
\frac{1}{n}
\frac{d(a_{\tau}^{n}-a_{\tau_0}^{n})}{d\tau}
\big(
-
\rho_{\tau_0}(0)
-
\rho_{\tau_0}(\infty)
+
\rho_{\tau_0}(0)
\big)
=
-\frac{1}{n}
\frac{d(a_{\tau}^{n}-a_{\tau_0}^{n})}{d\tau}
\rho_{\tau_0}(\infty)
=0,
\ee
\end{widetext}
if 
\be
\lim_{\mathtt{x}\to\infty}\rho_{\tau_0}(\mathtt{x})=0.
\ee
Here $a_{\tau_0}$ cannot be greater than $a_{\tau}$. It is simplest to work with $a_{\tau_0}=0$.

\section{Further properties  of ${\cal H}_0$}
\label{Sec5}

In formalisms of a Klauder type one usually works with coherent states and their resolutions of unity. We begin with eigenvectors of ${\cal H}_0$ and prove their completeness. Next, we rewrite ${\cal H}_0$ in terms of positions and canonical momenta. The latter form will be needed when it comes to matter fields and Hamiltonians of the form ${\cal H}={\cal H}_0+{\cal H}_1$.

\subsection{Eigenvectors}

The Hamiltonian
\be
{\cal H}_0=-\frac{1}{n}
\frac{\ell^n}{\mathtt x^n}x^{\mu}\, i\partial_{\mu}
\ee
is symmetric. 
Its eigenvectors are given by
\be
f_E(x)=f_E(0)e^{iE  \mathtt x^n/\ell^n},
\ee
for any $E$ (real or complex).
Note that for $\mathtt x^2=0$ we find $f_E(x)=f_E(0)$, so a nontrivial $f_E$ cannot vanish on the boundary $\partial V_+$.
$f_E(x)$ does not belong to our Hilbert space, which is not surprising.

\subsection{Completeness of the eigenvectors for real $E$}

In this subsection we set  $\ell=1$.
Let $x^{\mu}={\mathtt x}\,u^{\mu}$, $u^{\mu} u_{\mu}=1$. Both $x^{\mu}$ and $u^{\mu}$ are timelike and future-pointing. The formula
\be
d^nx
=
d\mathtt x\,\mathtt x^{n-1} du
=
d({\mathtt x}^n) \frac{d^{n-1}u}{n\sqrt{1+\bm u^2}}
\ee
defines an SO$(1,n-1)$-invariant  measure $du$, a natural curved-space generalization of $d^{n-1}x$.
Our well known quantum mechanics corresponds to $n=4$ and $\frac{d^{3}x}{\sqrt{1+\bm x^2/\mathtt x^2}}\approx d^{3}x$, an approximation valid for $\mathtt x$ of the order of the size of the observable universe and $\bm x$ achievable in present-day quantum measurements. 
The scalar product
\be
\langle f|g\rangle
&=&
\int_{V_+} d^nx \overline{f(x^0,\bm x)}g(x^0,\bm x)
\\
&=&
\int_0^\infty d\mathtt x\,\mathtt x^{n-1} \int_{u^2=1} du\,
\overline{f({\mathtt x}\,u^0,{\mathtt x}\bm u)}g({\mathtt x}\,u^0,{\mathtt x}\bm u),\nonumber
\ee
splits by means of the usual separation of variables into two scalar products:
\be
\langle A|A'\rangle_1
&=&
\int_0^\infty  d\mathtt x\,\mathtt x^{n-1} \overline{A(\mathtt x)}A'(\mathtt x)
\ee
and
\be
\langle B|B'\rangle_2
&=&
\int_{u^2=1} du\,\overline{B(\bm u)}B'(\bm u)\\
&=&
\int_{\mathbb{R}^{n-1}} \frac{d^{n-1}u}{\sqrt{1+\bm u^2}}\overline{B(\bm u)}B'(\bm u)
\ee
Let us thus consider some basis $B_j$ of special functions, orthonormal with respect to $\langle B_j|B_{j'}\rangle_2=\delta_{jj'}$, and define
\be 
\langle x|f_{E,j}\rangle=f_{E,j}(x)=e^{iE  \mathtt x^n}B_j(\bm u), \quad E\in\mathbb{R}.
\ee
Wave functions $g(x)$ can be nonzero only for $x\in V_+$, a condition preserved by $U_\tau$. Therefore,
\begin{widetext}
\be
\langle f_{E,j}|g\rangle
&=&
\int_{V_+} d^nx\overline{f_{E,j}(x)}g(x)
=
\int_{V_+} d^nx e^{-i  E  \mathtt x^n}\overline{B_j(\bm u)}g(\mathtt x u)\\
&=&
\int_0^\infty  d\mathtt x\, {\mathtt x}^{n-1} e^{-i  E  {\mathtt x}^n} \int_{u^2=1}du\overline{B_j(\bm u)}g(\mathtt x u)
\\
&=&
\frac{1}{n}\int_0^\infty  dt e^{-i  E  t} \int_{u^2=1}du\overline{B_j(\bm u)}g(t^{1/n} u)\quad \textrm{(here $t=\mathtt x^n$)}
\\
&=&
\frac{1}{n}\int_{-\infty}^\infty  dt e^{-i  E  t} \int_{u^2=1}du\overline{B_j(\bm u)}g\left(\sgn(t)|t|^{1/n} u\right)
\ee
The inverse Fourier transform,
\be
\frac{1}{2\pi}
\int_{-\infty}^\infty dE 
e^{i  E  t}
\langle f_{E,j}|g\rangle
&=&
\frac{1}{n}\int_{u^2=1}du\,\overline{B_j(\bm u)}g\left(\sgn(t)|t|^{1/n} u\right),
\ee
implies
\be
n\sum_jB_j(\bm u)\frac{1}{2\pi}
\int_{-\infty}^\infty dE 
e^{i  E  \mathtt x^n}
\langle f_{E,j}|g\rangle
&=&
\int_{u^2=1}du'\,\sum_jB_j(\bm u)\overline{B_j(\bm u')}g(\mathtt x u')=g(\mathtt x u),
\ee
\end{widetext}
or equivalently,
\be
\frac{n}{2\pi}
\int_{-\infty}^\infty dE 
\sum_j f_{E,j}(x)\langle f_{E,j}|g\rangle
=
g(x),
\ee
which can be written as the resolution of unity
\be
\frac{n}{2\pi}
\sum_j \int_{-\infty}^\infty dE 
|f_{E,j}\rangle\langle f_{E,j}|
=
\mathbb{I}.
\ee
We conclude that the spectrum of ${\cal H}_0$ consists of $\mathbb{R}$, and $f_{E,j}(x)$ form a complete set.
Various explicit forms of $B_j$ can be found in the literature that deals with quantum mechanics {\it on\/} Lobachevsky spaces.

\subsection{Cosmic four-position representation}

Dimensionless four-position representation is defined by:
\be
Q^\mu
&=&
\ell^{-1}x^\mu,\\
P_\mu
&=&
-i\ell \partial_{\mu},\\
V^\mu
&=&
\frac{1}{n}
\frac{\ell^{n-1}}{\mathtt x^n}x^{\mu},\\
{\cal H}_0
&=&
V^\mu P_\mu=P_\mu V^\mu .
\ee
The latter follows from the continuity equation $\partial_\mu V^\mu=0$.
The remaining basic commutators read:
\be
{[Q_\mu,P_\nu]} &=& ig_{\mu\nu},\\
{[Q_\mu,{\cal H}_0]}
&=&
i V_\mu.
\ee
Coupling of matter to space-time is given by
\be
{\cal H} &=&{\cal H}_0+{\cal H}_1\\
&=&
V^\mu P_\mu+{\cal H}_1(Q).\label{H0+H1}
\ee
In empty universe the wave function $\Psi_\tau(x)$ plays a role of a vacuum state. The space of such vacuum states is infinitely dimensional. The standard arguments leading to Ehrenfest's theorem in quantum mechanics are applicable here as well, so the average Minkowski-space position
\be
\langle x^\mu(\tau)\rangle = \int d^4x\, x^\mu|\Psi_\tau(x)|^2
\ee
defines a world-line of the center-of-mass of the empty universe. In a sense, the Copernican principle is spontaneously broken by the initial condition $\Psi_0(x)$.

Postponing the issue of matter fields to Sections~\ref{Sec int1}--\ref{Sec int4}, let us first concentrate on further properties of the empty universe, ${\cal H}_1(Q)=0$.

\section{Average size of the universe for $\Psi_\tau(x)$}
\label{Sec6}

Size of the universe is here described by the support properties of $\Psi_\tau(x)$. In our discussion we assume, for simplicity, that the support is given by a compact set, which is in fact somewhat too strong (we only need square integrability of $\Psi_\tau(x)$). Moreover, asymptotically for large $\tau$, the support gets concentrated in a neighborhood of a Milnean hyperboloid $x_{\mu}x^{\mu}=\mathtt x^2$, so consists of events that are approximately simultaneous from the point of view of $\tau$. Obviously, the support cannot be identified with the universe observable at $x^{\mu}$. The latter consists of the past cone of the event $x^{\mu}$.

Let us now investigate in more detail the timelike thickness of the wave packet for
\be
a_\tau=\ell \tau^{1/n} \le{\mathtt x}\le (\ell^n\tau +C)^{1/n}=b_\tau.
\ee
Denote
\be
\Delta_\tau &=&
(\ell^n\tau +C)^{1/n}-\ell \tau^{1/n}
\\
&=&
\frac{C}{\sum_{k=0}^{n-1}(\ell^n\tau +C)^{(n-1-k)/n}(\ell^n \tau)^{k/n}},\\
C
&=&
(\Delta_\tau+\ell \tau^{1/n})^n-\ell^n\tau.
\ee
$\lim_{\tau\to\infty} \Delta_\tau  = 0$ implies  that for large cosmic times the wave packet concentrates in a neighborhood of the hyperboloid $a_\tau^2=x_{\mu}x^{\mu}$. For $n=4$ the hyperboloid is given by
$\ell^2\sqrt{\tau}=c^2t^2-\bm x^2$. 
 
In our formalism, the four-dimensional volume $V^{(4)}$ of the universe is defined in a $\tau$-invariant way,
\be
V^{(4)} 
&=&
\int_{V_+} d^4 x |\Psi_\tau(x)|^2/\mathscr{Z}
\nonumber\\
&=&
\int_{V_+} d^4 x |\Psi_0(x)|^2/\mathscr{Z}=\Delta_\tau V_\tau^{(3)}=1/\mathscr{Z}.\label{V^{(4)}}
\ee
Writing $V_\tau^{(3)}\sim r_\tau^3$ we obtain a measure $r_\tau$ of the space-like size of the support of $\Psi_\tau(x)$, satisfying 
\be
\Delta_\tau r_\tau^3=\Delta_0 r_0^3.
\ee
Note that both $\Delta_\tau$ and  $r_\tau$ are invariant under the action of the Lorentz symmetry group of $\bar V_+$.
For $\tau_0=0$ the hyperboloid $\ell^2\sqrt{\tau}=c^2t^2-\bm x^2$ determines the gap, depicted in Fig.~\ref{Fig1}, between the support of $\Psi_\tau(x)$ and the light cone $c^2t^2-\bm x^2=0$. It is clear that $r_0$, in spite of being relativistically invariant, cannot be identified with  geodesic length computed along the light-cone, because the latter is always zero, while $\Delta_0=C^{1/4}$ is finite and nonzero, and thus $r_0$ is finite and nonzero as well.

Intuitively, $r_\tau$ represents a relativistically invariant average radius of the universe, an analogue of a half-width of a wave-packet. One has to keep in mind that, at $\tau=\tau_0=0$, the wave packet has a nontrivial  timelike profile, as illustrated by Fig.~\ref{Fig1}.

Let us now experiment with some estimates of the parameters involved in the construction. For example, take $\ell=1.61622837\times 10^{-35}$~m (Planck length), $t_H=4.55 \times 10^{17}$ s (Hubble time), 
$c=299\,792\,458$~m/s (velocity of light in vacuum), and define the quantum/cosmic Hubble time $\tau_H$ by $\ell\tau_H^{1/4} =ct_H$, 
\be
\tau_H &=&(ct_H/\ell)^4=5.07361\times 10^{243}.
\ee
For $n=4$,
\be
C
&=&
(\Delta_{\tau_H}+\ell \tau_H^{1/4})^4-\ell^4\tau_H
=
(\Delta_{\tau_H}+ct_H)^4-(ct_H)^4.\nonumber
\ee
Assuming $\Delta_{\tau_H} \approx \ell$, we arrive at the estimate
\be
C \approx 1.64081 \times 10^{44}~{\rm m}^4.
\ee
Initially, at $\tau_0=0$ the universe extends in timelike directions by approximately 1 AU, $\Delta_0=C^{1/4}\approx 1.13179\times 10^{11}$~m, that is by around 377 light seconds. Was our universe created in seven... minutes?

At the Hubble time we expect the universe to have the volume of the order of $(ct_H)^3$, hence the four-dimensional volume is of the order of $(ct_H)^3\ell$. Accordingly, we can estimate
\be
V^{(4)}= \Delta_0 V_0^{(3)}\sim (ct_H)^3\ell
\ee
The result is
\be
V_0^{(3)}\sim (ct_H)^3 \ell/\Delta_0\sim (ct_H)^3 \times 10^{-44}
\ee
The radius so defined changes with $\tau$ according to
\begin{widetext}
\be
r_\tau 
&=&
ct_H(\ell/\Delta_\tau)^{1/3}
\\
&=&
\left(
\frac{\ell (ct_H)^3}{(\ell+ct_H)^4-(ct_H)^4}
\right)^{1/3}
\Big(
(\ell^4\tau +C)^{3/4}
+
(\ell^4\tau +C)^{2/4}(\ell^4 \tau)^{1/4}
+
(\ell^4\tau +C)^{1/4}(\ell^4 \tau)^{2/4}
+
(\ell^4 \tau)^{3/4}
\Big)^{1/3}.
\ee
The hyperboloid formula $\ell^4 \tau=(x_{\mu}x^{\mu})^2={\mathtt x}_\tau^4$ leads to
\be
r_\tau 
&=&
\left(
\frac{\ell (ct_H)^3}{(\ell+ct_H)^4-(ct_H)^4}
\right)^{1/3}
\Big(
({\mathtt x}_\tau^4 +C)^{3/4}
+
({\mathtt x}_\tau^4 +C)^{2/4}{\mathtt x}_\tau
+
({\mathtt x}_\tau^4 +C)^{1/4}{\mathtt x}_\tau^2
+
{\mathtt x}_\tau^3
\Big)^{1/3}\nonumber\\
&\approx&
4^{-1/3}
\Big(
({\mathtt x}_\tau^4 +C)^{3/4}
+
({\mathtt x}_\tau^4 +C)^{2/4}{\mathtt x}_\tau
+
({\mathtt x}_\tau^4 +C)^{1/4}{\mathtt x}_\tau^2
+
{\mathtt x}_\tau^3
\Big)^{1/3}.
\label{r_tau}
\ee
\end{widetext}
Here, ${\mathtt x}_\tau$ defines the hyperboloid that restricts the time-like extent of the support from below (that is, ${\mathtt x}_\tau$ measures the space-time gap between the support and $\partial V_+$).

For large ${\mathtt x}_\tau$, say ${\mathtt x}_\tau=\ell \tau_H^{1/4}$, one finds an approximately linear relation between 
$r_\tau$ and ${\mathtt x}_\tau$,
\be
r_\tau 
\approx
{\mathtt x}_\tau=\ell \tau^{1/4}.
\ee
Let us stress again that estimates such as as (\ref{r_tau}) deal with the support of $\Psi_\tau(x)$, so they effectively determine the volume one often encounters in quantum optics in `finite-box' mode decomposition of fields. In our model the volume is finite but  its size grows with  ${\mathtt x}_\tau$ approximately linearly, that is, proportionally to the fourth root of the cosmic/quantum time. 

It is clear that such a quantization volume has nothing to do with the observable universe that should be identified with the past cone of the argument $x^{\mu}$ in $\Psi_\tau(x)$. The observable universe has here the same meaning as in the Milnean cosmology. 

\begin{widetext}
\section{Gap hyperboloid, $\ell^2\sqrt{\tau}=c^2t^2-\bm x^2$}
\label{Sec7}

The gap hyperboloid may be regarded as a semiclassical characteristic of the universe. 

With the cosmic-time parametrization of $V_+$,
\be
(x^0,x^1,x^2,x^3)
=
\ell \tau^{1/4}(\cosh \Xi,\sinh \Xi \cos\Phi\sin\theta,\sinh \Xi \sin\Phi\sin\theta,\sinh \Xi \cos\theta),
\ee
the Minkowski-space metric $g_{\mu\nu}$ of $V_+$ can be rewritten in terms of $(\tilde x^0,\tilde x^1,\tilde x^2,\tilde x^3)=(\tau,\Xi,\Phi,\theta)$, 
\be
ds^2 
&=&
g_{\mu\nu}dx^{\mu} dx^{\nu}
\\
&=&
(\ell d\tau^{1/4})^2
-
(\ell \tau^{1/4})^2
\left(
d\Xi^2
+
\sinh^2 \Xi \sin^2\theta \,d\Phi^2
+
\sinh^2 \Xi
\,d\theta^2
\right)\label{103}\\
&=&
\frac{1}{16}\ell^2 \tau^{-3/2} d\tau^2
-
\ell^2 \tau^{1/2}
\left(
d\Xi^2
+
\sinh^2 \Xi \sin^2\theta \,d\Phi^2
+
\sinh^2 \Xi
\,d\theta^2
\right).\label{104}
\ee
The form (\ref{103}) is the standard Milnean metric, provided one threats $\ell\tau^{1/4}/c$ as the standard (classical) cosmological time (not to be confused with $\tau$ itself, our  quantum cosmic time parameter). The corresponding Hubble diagram for distance modulus vs. redshift is known to agree with the observed expansion of our universe  \cite{NGS,RM}.
On the other hand, the form (\ref{104}) shows that for the present values of $\tau$ (i.e. $\tau_H\sim 10^{243}$), the timelike component of the metric is a very tiny number,
\be
g_{\tau\tau} = \frac{1}{16}\ell^2 \tau^{-3/2}\sim 10^{-435}\textrm{m}^2\label{105},
\ee
as if the background space-time was effectively 3-dimensional. The latter agrees with the support properties of $\Psi_\tau(x)$ because $\Psi_{\tau_H}(x)\neq 0$ only in a narrow future-neighborhood of the gap hyperboloid. At the other extreme is the case of $\tau\approx 0$, where $g_{\tau\tau}$ is large in comparison to
\be
g_{\Xi\Xi}\approx g_{\Phi\Phi}\approx g_{\theta\theta}\approx 0,\label{106},
\ee
as if the space-time was 1-dimensional and consisted of time only. 

Of course, the estimates (\ref{105}) and (\ref{106}), reflect asymptotic properties of the metric tensor of the background space-time and not of the universe itself, identified here with the set of those points $x$ that satisfy $\Psi_\tau(x)\neq 0$. However, this set is partly characterized by properties of the gap hyperboloid, which in turn is characterized by the evolution parameter $\tau$. The asymptotic properties of (\ref{104}) agree with the intuitive classical picture of the universe that evolves from a single point at $\tau=0$ into a 3-dimensional space as $\tau\to\infty$.

An exact relation between Minkowskian space-time $x^0=ct$ and the cosmic/quantum $\tau$ is implied by the hyperboloid equation
$\ell^2\sqrt{\tau}=c^2t^2-\bm x^2$, so 
\be
ct(\tau,\bm x)
&=&
\sqrt{\ell^2\sqrt{\tau}+\bm x^2}.
\ee
In consequence,
\be
ct(\tau+\Delta\tau,\bm x)
&=&
\sqrt{\ell^2\sqrt{\tau+\Delta\tau}+\bm x^2}
=
ct(\tau,\bm x)
+
\frac{\ell^2}
{4\sqrt{\tau}\sqrt{\ell^2\sqrt{\tau}+\bm x^2}}
\Delta\tau+\dots\\
&=&
ct(\tau,\bm x)
+
\frac{\ell}
{4\tau^{3/4}\sqrt{1+\left(\frac{\bm x}{\ell\sqrt[4]{\tau}}\right)^2}}
\Delta\tau+\dots\label{99}
\ee
Assuming that present-day observers deal with cosmic times of the order of the Hubble time, $\ell \tau^{1/4}\approx ct_H$,
and systems whose sizes are negligible in comparison to the size of the universe, $\bm x/(ct_H)\approx 0$, we can neglect the square root occurring in the denominator of (\ref{99}), 
\be
ct(\tau+\Delta\tau,\bm x)
&=&
ct(\tau,\bm x)+c\Delta t(\tau,\bm x)
\approx
ct(\tau,\bm x)
+
\frac{\ell^4}
{4(ct_H)^3}
\Delta\tau.
\ee
\end{widetext}
The usual $\Delta t$ we encounter in elementary undergraduate nonrelativistic definition of velocity or acceleration is related to our $\Delta \tau$ by 
\be
c\Delta t=\frac{\ell^4}
{4(ct_H)^3}
\Delta\tau.
\ee

It is intriguing that Wiener, in his MIT lectures on Brownian motion \cite{Wiener}, introduced the notion of a roughness of a curve (measuring straightness of a string $x(s)$ passing through a given sequence of holes) by 
\be
\max \frac{x(s+\tau)-x(s)}{\tau^{1/4}}.
\ee
Wiener's `roughness' thus resembles a derivative of $x(\tau)$ but computed with respect to $t(\tau)$ and not just $\tau$ itself. 

\section{`Average radius of the universe' vs.  spacelike geodesic distance}
\label{Sec8}

Let us continue with $n=1+3$. Consider $x^{\mu}$ and $y^{\mu}$ that belong to the same hyperboloid, $\mathtt x^2=x_{\mu}x^{\mu}=y_{\mu}y^{\mu}=\mathtt y^2$. Define 
$(x\cdot y)/\mathtt x^2=\cosh \xi$ (here $x\cdot y=x_{\mu}y^{\mu}$). Then
\be
\xi
&=&
\left\{
\begin{array}{r}
\ln \Big((x\cdot y)/\mathtt x^2+\sqrt{(x\cdot y)^2/\mathtt x^4-1}\Big),\quad \textrm{if $\xi\geq 0$},\\
-\ln \Big((x\cdot y)/\mathtt x^2+\sqrt{(x\cdot y)^2/\mathtt x^4-1}\Big),\quad \textrm{if $\xi\leq 0$},
\end{array}
\right.
\ee
so the geodesic distance between $x$ and $y$, computed along the hyperboloid,  is
\be
{\mathtt x}|\xi|
&=&
{\mathtt x}\ln \frac{x\cdot y+\sqrt{(x\cdot y)^2-\mathtt x^4}}{\mathtt x^2}
\ee
With $\mathtt x\to\infty$ the geodesic distance is just the Euclidean distance in $\mathbb{R}^3$. 

Writing ${\mathtt x}|\xi|=r$,  we can parametrize Lorentz transformations mapping $y^{\mu}$ into $x^{\mu}$ by means of $r$, the distance between the two points. Taking ${\mathtt x}_\tau=\ell\tau^{1/4}$, $y^{\bm a}=({\mathtt x}_\tau,\bm 0)$, $x^{\bm a}=(\sqrt{{\mathtt x}_\tau^2+\bm x^2},\bm x)$, we find
\be
(x\cdot y)/\mathtt x_\tau^2=x_0/\mathtt x_\tau=\cosh (r_\tau/\mathtt x_\tau).
\ee
For any unit 3-vector $\bm n$ we conclude that
\be
(x_0,\bm x)
&=&
{\mathtt x}_\tau\big(\cosh (r_\tau/\mathtt x_\tau),\bm n\sinh (r_\tau/\mathtt x_\tau)\big),\label{r geo}
\ee
is located  in geodesic distance $r_\tau$ from the origin $\bm x=\bm 0$. The distance is computed along the hyperboloid $\ell^2\sqrt{\tau}=x_{\mu}x^{\mu}={\mathtt x}_\tau^2$. The result is Lorentz invariant, so is typical of any choice of the origin.

\begin{figure}
\includegraphics[width=8 cm]{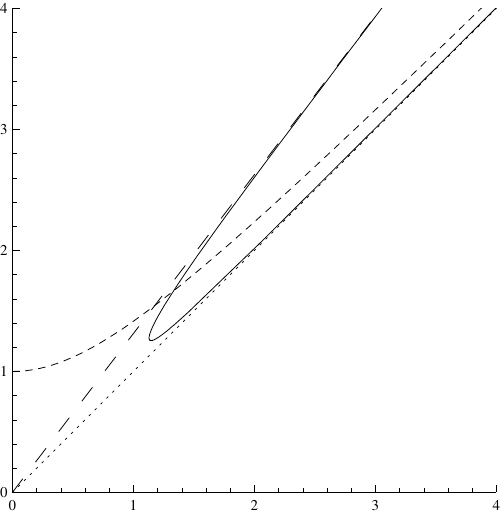}
\caption{The hyperboloid $x_{\mu}x^{\mu}=\Delta_0^2$, in units where $\Delta_0=1$ (dashed) and the light-cone (dotted). The full line represents the world line (\ref{r geo}), plotted in the half-plane spanned by $(1,\bm 0)$ and $(0,\bm n)$. When $\tau$ tends to 0, the curve escapes toward future-null infinity. The wide-dashed straight line is the asymptote (\ref{asymptota}), 
$\tau\mapsto {\mathtt x}_\tau\big(\cosh 1,\bm n\sinh 1\big)$}
\label{Fig3}
\end{figure}

Let us now consider points separated by geodesic distance  $r_\tau$  on hyperboloid $\mathtt x_\tau$, but assume that the distance coincides with the `average radius of the universe' $r_\tau$ given by  (\ref{r_tau}),
\begin{widetext}
\be
r_\tau /\mathtt x_\tau
&\approx&
4^{-1/3}
\Big(
({\mathtt x}_\tau^4 +C)^{3/4}
+
({\mathtt x}_\tau^4 +C)^{2/4}{\mathtt x}_\tau
+
({\mathtt x}_\tau^4 +C)^{1/4}{\mathtt x}_\tau^2
+
{\mathtt x}_\tau^3
\Big)^{1/3}/\mathtt x_\tau.
\ee
\end{widetext}
Asymptotically, $\lim_{\tau\to 0}r_\tau /\mathtt x_\tau=\infty$, $\lim_{\tau\to \infty}r_\tau /\mathtt x_\tau=1$.
For large $\tau$ the world-vectors whose geodesic distance from  $\bm x=\bm 0$ equals $r_\tau\approx {\mathtt x}_\tau=\ell\tau^{1/4}$
are located on straight and time-like world-lines (Fig.~\ref{Fig3})
\be
\tau\mapsto
{\mathtt x}_\tau\big(\cosh 1,\bm n\sinh 1\big)
=
\ell\tau^{1/4}
(1.543, 1.175\, \bm n).\label{asymptota}
\ee
Such world lines may be regarded as quantum analogues of generators of an expanding boundary of our universe. Interestingly, asymptotically, for late cosmic times the resulting boundary does not expand with velocity of light, but rather with $v=c\tanh 1\approx 0.76\,c$. It is intriguing that 
$\tanh 1$ is a neutral element of multiplication in the arithmetic of relativistic velocities.

\section{Empty universe as a reservoir for matter fields}
\label{Sec int1}
So far, our universe is empty. Matter fields should be included by means of (\ref{H0+H1}). Leaving a detailed discussion of explicit examples to a separate paper let us outline the construction of ${\cal H}_1(Q)$. 

We are guided by the asymptotic correspondence principle with ordinary quantum mechanics and quantum field theory in our part of the universe, formulated as follows.

We assume that after some 13-14 billion years of the cosmic evolution the matter fields within our Galaxy should evolve by means of a matter-field Hamiltonian
\be
{\cal H}_1({\tau_H})
&\approx&
\mathscr{Z}
\int_{V^{(3)}_{\tau_H}}
d^3x\, T_{00}(ct_H,\bm{x}).
\ee
Here $V^{(3)}_{\tau_H}$ is the effective volume of the universe at $\tau=\tau_H$, as implied by   Eq. (\ref{V^{(4)}}),
$\mathscr{Z}$ is a renormalization constant, and $T_{\mu\nu}$ is an energy-momentum tensor of some matter field.

Introducing the characteristic function $\chi_{{V^{(3)}_{\tau_H}}}(\bm{x})$ of ${V^{(3)}_{\tau_H}}$,
as well as the approximation of the measure,
\be
d^{3}x
\approx
\frac{d^{3}x}{\sqrt{1+\bm x^2/(ct_H)^2}}
\ee
we can write
\be
{\cal H}_1({\tau_H})
&\approx&
\mathscr{Z}
\int_{\mathbb{R}^3}
\frac{d^{3}x}{\sqrt{1+\frac{\bm x^2}{(ct_H)^2}}}
\chi_{{V^{(3)}_{\tau_H}}}(\bm{x})
T_{00}(ct_H,\bm{x})\nonumber\\
&=&
\mathscr{Z}
\int_{V_+}
d^{4}x\,\delta\big(\mathtt{x}^2-(ct_H)^2\big)
\chi_{{V^{(3)}_{\tau_H}}}(\bm{x})
T_{00}(x).
\label{120}
\ee
Constant $\mathscr{Z}$  should be chosen so that
\be
\mathscr{Z}
\int_{\mathbb{R}^3}
\frac{d^{3}x}{\sqrt{1+\frac{\bm x^2}{(ct_H)^2}}}
\chi_{{V^{(3)}_{\tau_H}}}(\bm{x})=1.\label{121}
\ee
Equivalently,
\be
1
&=&
\mathscr{Z}
\int_{V_+}
d^{4}x\,\delta\big(\mathtt{x}^2-(ct_H)^2\big)
\chi_{{V^{(3)}_{\tau_H}}}(\bm{x})
\nonumber\\
&=&
\int_{V_+}
d^{4}x\,|\Psi_{\tau_H}(x)|^2
=
\int_{V_+}
d^{4}x\,|\Psi_\tau(x)|^2.\label{122}
\ee
Comparison of formulas (\ref{120})--(\ref{122}) leads to the conclusion that an  exact expression, valid for all $\tau$, should read
\be
{\cal H}_1(\tau)
&=&
\int_{V_+}
d^{4}x\,|\Psi_\tau(x)|^2
x^\mu x^\nu T_{\mu\nu}(x)/\mathtt{x}^2\\
&=&
\int_{V_+}
d^{4}x\,|\Psi_\tau(x)|^2
{\cal H}_1(x)\label{124'}\\
&=&
\int_{V_+}
d^{4}x\,\langle\Psi_0|U_\tau^\dag |x\rangle\langle x|U_\tau |\Psi_0\rangle
{\cal H}_1(x)
\label{123}
\ee
where we have used the fact that $x^\mu/(ct_H)\approx (1,\bm 0)$ for $x^\mu$ in a small (say, Galaxy-scale) neighborhood of our labs.

One concludes that what we regard as a total Hamiltonian that governs time evolution of matter in the present-day and our-part quantum universe looks like a partial average over the reservoir of an {\it interaction-picture\/} Hamiltonian
\be
{\cal H}_1(Q_\tau)
&=&
\int_{V_+}
d^{4}x\,U_\tau^\dag |x\rangle\langle x|U_\tau \otimes
{\cal H}_1(x).
\label{125}
\ee
The full Schr\"odinger-picture  Hamiltonian thus reads
\be
{\cal H}
&=&
{\cal H}_0+{\cal H}_1\\
&=&
V^\mu P_\mu\otimes \mathbb{I}
+
\int_{V_+}
d^{4}x\,|x\rangle\langle x| \otimes
{\cal H}_1(x)\label{total H}\\
&=&
V^\mu P_\mu\otimes \mathbb{I}
+
{\cal H}_1(Q).\label{total H'}
\ee
The presence of $|\Psi_\tau(x)|^2$ in (\ref{124'}) can be also interpreted by means of a certain weak limit $N\to\infty$, if one replaces in (\ref{total H}) the single projector $I(x,1)=|x\rangle\langle x|$ by the frequency-of-success operator
\be
I(x,N) &=&
\frac{1}{N}\big(
\underbrace{I(x,1)\otimes \mathbb{I}\otimes\dots\otimes\mathbb{I}}_N+
\dots
+
\mathbb{I}\otimes\dots\otimes\mathbb{I}\otimes I(x,1)
\big)\label{130_}
\ee
employed in weak quantum laws of large numbers \cite{LLN1,LLN2,LLN3,LLN4}. Operator (\ref{130_}) occurs also in commutators of field operators if fields are quantized by means of reducible representations of the oscillator algebra \cite{MC1,MC2,MC3,MC4,MC5}. 
The free part then takes the form of a free $N$-particle bosonic extension of  ${\cal H}_0(1)=V^\mu P_\mu$, i.e. 
\be
{\cal H}_0(N)
&=&
{\cal H}_0(1)\otimes \mathbb{I}\otimes\dots\otimes\mathbb{I}+
\dots
+
\mathbb{I}\otimes\dots\otimes\mathbb{I}\otimes
{\cal H}_0(1).
\ee
It is then a standard result that at the level of matrix elements the limit $N\to\infty$ is equivalent to the replacement $I(x,N)\to |\Psi_\tau(x)|^2$, where $|\Psi_\tau\rangle$ is interpreted as a vacuum state, which agrees with the intuition that cosmological vacuum corresponds to an empty universe. Moreover, parameters such as $\mathscr{Z}$, related to 
$|\Psi_\tau(x)|^2$ by formulas (\ref{120})--(\ref{122}), can be shown to play the role of renormalization constants in exactly the same sense as the one employed in quantum field theory. 

Accordingly, operators of the form (\ref{124'}) will occur as weak limits $N\to\infty$ of
\be
{\cal H}(N)={\cal H}_0(N)\otimes\mathbb{I}+
\int_{V_+}
d^{4}x\,I(x,N) \otimes
{\cal H}_1(x)\label{132_},
\ee
if the limit is taken in the interaction picture.
Hamiltonian (\ref{132_}) for a finite $N$ describes an $N$-point universe, an analogue of an $N$-particle state, where each of the particles is pointlike and bosonic. For large $N$ the universe becomes a Bose-Einstein condensate of pointlike objects, whose probability density in space-time is given by $|\Psi_\tau(x)|^2$. Let us stress that these pointlike entities should not be treated as matter-field particles, but as points of the universe itself.
For various technicalities of the weak limits  see 
 \cite{LLN1,LLN2,LLN3,LLN4,MC1,MC2,MC3,MC4,MC5}, but a detailed exposition of the approach is beyond the scope of the present paper. The model which is formally closest to what we encounter here is the case of a classical current interacting with quantized electromagnetic field, discussed in detail in \cite{MC4}. Readers interested in generalization based on (\ref{132_}) should first understand the construction from \cite{MC4}.

Let us note that the choice 
\be
{\cal H}_1(x)
&=&
x^\mu x^\nu T_{\mu\nu}(x)/\mathtt{x}^2
\ee
is motivated by isotropy, uniformity, manifest covariance and, first of all, the correspondence principle with $T_{00}$ for large $\tau\approx \tau_H$. We do not need the usual argument based on continuity equation $\nabla^a T_{ab}=0$, because 
(\ref{total H}) is anyway independent of $x_0$ and $\tau$. This is why the issues such as non-vanishing trace of $T_{ab}$ or transvection of $T_{ab}$ with Killing fields are irrelevant in this context. The transvection with $u^\mu=x^\mu/\mathtt{x}$ can be postulated  regardless of its property of being or not being a Killing field of some symmetry.

Schr\"odinger-picture  Hamiltonian describes evolution of the entire `universe plus matter' system. Average energy of the whole system is conserved but, of course, the energy of matter alone is not conserved. However, at large $\tau$ the averaged-over-reservoir matter Hamiltonian is essentially the standard Hamiltonian but evaluated in a finite and growing with time `quantization volume'. 

The structure of the Hilbert space associated with ${\cal H}_1$ also resembles the one occurring in the `quantum time' formalism of Page and Wooters \cite{PW,GLM}.

\section{Effective conformal coupling of matter and geometry} 
\label{Sec int2}
The universe is defined in terms of $|\Psi_\tau(x)|^2$ which effectively determines coupling of matter and space-time by means of the formula for the averaged-over-reservoir interaction Hamiltonian (or the weak large-number limit of (\ref{132_}))
\be
{\cal H}_1(\tau)
&=&
\int_{V_+}
d^{4}x\,|\Psi_\tau(x)|^2
{\cal H}_1(x).\label{129}
\ee
There are two natural ways of interpreting (\ref{129}) as a representation of coupling between matter and geometry. 

The first one is based on the identification 
\be
|\Psi_\tau(x)|^2=\sqrt{|g_\tau(x)|}\label{det g}
\ee
In 4-dimensional background Minkowski space $\cal M$ with metric $g_{\mu\nu}$ we can write \cite{PR,P1983}
\be
g_{\tau\mu\nu}(x)=|\Omega_\tau(x)|^2 g_{\mu\nu},
\ee
so that
\be
g_\tau(x)
=
-
|\Omega_\tau(x)|^8
\ee
and $|\Psi_\tau(x)|^2=|\Omega_\tau(x)|^4$. Here we again have two options. Firstly,  we can employ the usual strategy and demand that $\Omega_\tau(x)$ be real and non-negative, hence
\be
\Omega_\tau(x)
&=&
|\Psi_\tau(x)|^{1/2},\label{136,}\\
g_{\tau\mu\nu}(x)
&=&
|\Psi_\tau(x)| g_{\mu\nu},\\
g_\tau^{\mu\nu}(x)
&=&
|\Psi_\tau(x)|^{-1} g^{\mu\nu}.\label{138,}
\ee
Recall that the universe is identified with $x\in\cal M$ fulfilling $\Psi_\tau(x)\neq 0$.

Secondly, we can write
\be
\overline{\Psi_\tau(x)}\Psi_\tau(x)=\overline{\Omega_\tau(x)^2}\Omega_\tau(x)^2,
\ee
so that 
\be
\Omega_\tau(x)=\Psi_\tau(x)^{1/2}\label{137}
\ee
is complex. We know that complex $\Omega_\tau(x)$ will lead to a connection with torsion \cite{P1983}. 

However, for Hamiltonian densities
\be
{\cal H}_1(x)={\cal H}_1\big(\phi_{\cal A}(x),\nabla\phi_{\cal A}(x)\big)
\ee
which are quadratic in matter fields $\phi_{\cal A}(x)$, there exists yet another theoretical possibility. Namely,
we can demand
\be
{\cal H}_1(\tau)
&=&
\int_{V_+}
d^{4}x\,|\Psi_\tau(x)|^2
{\cal H}_1\big(\phi_{\cal A}(x),\nabla\phi_{\cal A}(x)\big)\label{138}\\
&=&
\int_{V_+}
d^{4}x\,
{\cal H}_1\big[\Psi_\tau\phi_{\cal A}(x),D(\Psi_\tau\phi_{\cal A}(x))\big],\label{139}
\ee
where $D$ and $\nabla$ are spinor covariant derivatives related by \cite{PR,P1983}
\be
\nabla_\mu\hat\ve(x)_{BC}=0,\quad D_\mu\Psi_\tau(x)\hat\ve(x)_{BC}=0.\label{nabla D}
\ee
Spinor $\hat\ve(x)_{BC}$ is unspecified as yet. 
Typically either $\nabla$ or $D$ has non-vanishing torsion. (\ref{139}) suggests that $D=\partial$ is the usual flat torsion-free covariant derivative in Minkowski space, and thus
\be
\Psi_\tau(x)\hat\ve(x)_{BC}=\ve_{BC}
\ee
is the usual flat Minkowski-space `metric' spinor. Hence,
\be
\hat\ve(x)_{BC} &=& \Psi_\tau(x)^{-1}\ve_{BC}=\Omega_\tau(x)\ve_{BC},\label{143}\\
\hat\ve(x)^{BC} &=& \Psi_\tau(x)\ve^{BC}=\Omega_\tau(x)^{-1}\ve^{BC},\\
\hat\ve(x)_{B'C'} &=& \overline{\Psi_\tau(x)^{-1}}\ve_{B'C'}=\bar\Omega_\tau(x)\ve_{B'C'},\\
\hat\ve(x)^{B'C'} &=& \overline{\Psi_\tau(x)}\ve^{B'C'}=\bar\Omega_\tau(x)^{-1}\ve^{B'C'}.\label{146}
\ee
We have skipped $\sqrt{|g|}$ in (\ref{138})--(\ref{139}) because now the conformal transformations 
are not regarded as changes of coordinates on the same space-time, but as modifications of the space-time itself. The lack of square root in (\ref{143})--(\ref{146}) is not a typographic error. The construction leading to (\ref{143})--(\ref{146}) is not equivalent to the one that has led to (\ref{137}).

$\nabla$ and $D$ can be chosen in many different ways. The standard paradigm is to assume conformal invariance of field equations satisfied by matter fields (which excludes massive fields), and demand that connections be torsion-free. However, a complex conformal transformation naturally introduces non-vanishing torsion. Moreover, the formalism should not crucially depend on mass of matter fields. In what follows we discuss a conection which has the required properties.  

Our discussion is based on the Minkowski-space background. However, the same analysis can be performed in space-times that are conformally flat, which includes FRW cosmologies \cite{Conf1,Conf2,Conf3}.

\section{Conformal covariance of arbitrary-mass matter fields}
\label{Sec int3}
Typically conformal covariance is associated with massless fields or twistors \cite{PR}. In the present section we will take a closer look at the standard construction of spinor covariant derivative, leading us to a simple form of connection that does not require $m=0$ for  conformal covariance of matter fields. We switch to the standard spinor notation with space-time abstract indices written in lowercase Roman fonts (in the previous sections we avoided formulas of the form $x^a$ because $a$ could be confused with the scale factor). The conformal factor $\Omega$ can denote either $\Psi_\tau(x)^{1/2}$ or $\Psi_\tau(x)^{-1}$.
By $\partial_a=\partial_{AA'}$ we denote the flat torsion-free covariant derivative in 4-dimensional Minkowski space with signature $(+,-,-,-)$. $g_{ab}$ is the Minkowski-space metric.

We begin with
\be
\nabla_a \Om\ve_{BC} &=&0,\label{300}\\
\nabla_a \Om^{-1}\ve^{BC} &=&0,\\
\nabla_a \bar\Om\ve_{B'C'} &=&0,\\
\nabla_a \bar\Om^{-1}\ve^{B'C'} &=&0.\label{303}
\ee
When comparing our formulas with Eq. (5.6.11) in \cite{PR}, keep in mind that our $\partial_a$ is denoted in the Penrose-Rindler monograph by $\nabla_a$, so our $\nabla_a$ stands for their $\hat\nabla_a$. Practically, the only consequence of this conflict of notation is in the opposite sign of the torsion tensor.

Spinor connection is denoted by
\be
\nabla{_{a}}f{_{B}}
&=&
\partial{_{a}}f{_{B}}
-
\Theta{_{a}}{_{B}}{^{C}}f{_{C}}\\
\nabla{_{a}}f{_{B'}}
&=&
\partial{_{a}}f{_{B'}}
-
\bar\Theta{_{a}}{_{B'}}{^{C'}}f{_{C'}}.
\ee
Equations (\ref{300})--(\ref{303}) imply
\be
\Theta{_{a}}{_{B}}{_{C}}
&=&
\Theta{_{a}}{_{(B}}{_{C)}}
+\frac{1}{2}
\Om^{-1}\partial{_{a}}\Omega\,
\ve{_{B}}{_{C}},\label{151}\\
\bar\Theta{_{a}}{_{B'}}{_{C'}}
&=&
\bar\Theta{_{a}}{_{(B'}}{_{C')}}
+\frac{1}{2}
\bar\Om^{-1}\partial{_{a}}\bar\Omega\,
\ve{_{B'}}{_{C'}}.\label{152}
\ee
The torsion tensor is given by
\begin{widetext}
\be
(\nabla{_{a}}\nabla{_{b}}-\nabla{_{b}}\nabla{_{a}})f
&=&
\Big(
-
\Theta{_{a}}{_{B}}{_{C}}\ve{_{B'}}{_{C'}}
-
\bar\Theta{_{a}}{_{B'}}{_{C'}}\ve{_{B}}{_{C}}
+
\Theta{_{b}}{_{A}}{_{C}}\ve{_{A'}}{_{C'}}
+
\bar\Theta{_{b}}{_{A'}}{_{C'}}\ve{_{A}}{_{C}}
\Big)
\nabla{^{c}}f\\
&=&
T{_a}{_b}{_c}\nabla{^{c}}f
\ee
\end{widetext}
 (note the sign difference with respect to Eq. (4.4.37) in \cite{PR}). 
The study of complex conformal transformations was initiated in \cite{P1983} with the conclusion that 
$T{_a}{_b}{_c}\neq 0$ may be an interesting alternative to the usual choice of $T{_a}{_b}{_c}=0$. 
Although we generally agree here with Penrose, we will not exactly follow the suggestions form \cite{P1983}. However, before we present our own preferred spinor connection let us first recall the results from \cite{P1983}.

\subsection{The case $T_{abc}=0$}
\label{Sec A}

Assume that $T_{abc}=0$ in addition to (\ref{300})--(\ref{303}). Then (cf. (4.4.47) in \cite{PR})
\be
\Theta{_{a}}{_{B}}{^{C}}
&=&
i
\Pi{_{a}}\ve{_{B}}{^{C}}
+
\Upsilon{_{B}}{_{A'}}\ve{_{A}}{^{C}}\label{157}\\
&=&
\frac{1}{4}\Big(\Omega^{-1}\partial{_{a}}\Omega-\bar\Omega^{-1}\partial{_{a}}\bar\Omega\Big)
\ve{_{B}}{^{C}}
+
\ve{_{A}}{^{C}}\partial{_{B}}{_{A'}}\ln|\Omega|,
\\
\bar\Theta{_{a}}{_{B'}}{^{C'}}
&=&
-i
\Pi{_{a}}\ve{_{B'}}{^{C'}}
+
\Upsilon{_{A}}{_{B'}}\ve{_{A'}}{^{C'}}\label{158}\\
&=&
-\frac{1}{4}\Big(\Omega^{-1}\partial{_{a}}\Omega-\bar\Omega^{-1}\partial{_{a}}\bar\Omega\Big)
\ve{_{B'}}{^{C'}}
+
\ve{_{A'}}{^{C'}}\partial{_{A}}{_{B'}}\ln|\Omega|.
\ee
The world-vectors $\Pi_a$ and $\Upsilon_a$ are real.
The particular case $\Omega=e^{i\theta}$, $|\Omega|=1$, leads to
\be
\Theta{_{a}}{_{B}}{_{C}}
=
\frac{i}{2}\partial_a\theta\, \ve{_{B}}{_{C}}\label{IvW}
\ee
and was discussed by Infeld and van der Waerden in their attempt of deriving electromagnetic fields directly from spinor connections \cite{IvW}. Connection (\ref{IvW}) bears a superficial similarity to the antisymmetric connection we discuss in Sec.~\ref{Sub C}.  However, the essential difference between (\ref{IvW}) and (\ref{174}) is that the latter can be real.

Transformation
\be
\Phi{_{B_1}}{_{B_2}}{_{\dots}}{_{B_n}}
&=&
\Omega^{\frac{n}{4}-\frac{1}{2}}\bar\Omega^{-\frac{n}{4}-\frac{1}{2}}
\varphi{_{B_1}}{_{B_2}}{_{\dots}}{_{B_n}}\label{161}
\ee
implies
\begin{widetext}
\be
\Omega^{-\frac{n}{4}+\frac{1}{2}}\bar\Omega^{\frac{n}{4}+\frac{1}{2}}
\nabla{_{a}}\Phi{_{B_1}}{_{B_2}}{_{\dots}}{_{B_n}}
&=&
\partial{_{a}}
\varphi{_{B_1}}{_{B_2}}{_{\dots}}{_{B_n}}
\label{340}\\
&\pp=&
-
(\partial{_{a}}\ln|\Omega|)
\varphi{_{B_1}}{_{B_2}}{_{\dots}}{_{B_n}}
-
(\partial{_{B_1}}{_{A'}}\ln|\Omega|)
\varphi{_{A}}{_{B_2}}{_{\dots}}{_{B_n}}
-
\dots
-
(\partial{_{B_n}}{_{A'}}\ln|\Omega|)
\varphi{_{B_1}}{_{B_2}}{_{\dots}}{_{A}}
\nonumber
\ee
If $\varphi{_{B_1}}{_{\dots}}{_{B_n}}$ is totally symmetric then
\be
-
(\partial{_{A}}{_{A'}}\ln|\Omega|)
\varphi{_{B_1}}{_{B_2}}{_{\dots}}{_{B_n}}
-
(\partial{_{B_1}}{_{A'}}\ln|\Omega|)
\varphi{_{A}}{_{B_2}}{_{\dots}}{_{B_n}}
-
\dots
-
(\partial{_{B_n}}{_{A'}}\ln|\Omega|)
\varphi{_{B_1}}{_{B_2}}{_{\dots}}{_{A}}
\nonumber
\ee
is totally symmetric in $A{{B_1}}{{\dots}}{{B_n}}$. Transvecting $A$ with any  $B_j$ we obtain a conformally covariant formula
\be
\Omega^{-\frac{n}{4}+\frac{1}{2}}\bar\Omega^{\frac{n}{4}+\frac{1}{2}}
\nabla{^{B_j}}{_{A'}}\Phi{_{B_1}}{_{\dots}}{_{B_j}}{_{\dots}}{_{B_n}}
=
\partial{^{B_j}}{_{A'}}
\varphi{_{B_1}}{_{\dots}}{_{B_j}}{_{\dots}}{_{B_n}}.
\ee
\end{widetext}
The massless-field equation 
\be
\partial{^{B_j}}{_{A'}}
\varphi{_{B_1}}{_{\dots}}{_{B_j}}{_{\dots}}{_{B_n}}
=0
\ee
thus implies 
\be
\nabla{^{B_j}}{_{A'}}\Phi{_{B_1}}{_{\dots}}{_{B_j}}{_{\dots}}{_{B_n}}
=0.\label{standard konf}
\ee
Conformal transformation (\ref{161}) was introduced in \cite{P1983}. 
If $\Omega=|\Omega|$ then (\ref{161}) takes the well known form
\be
\Phi{_{B_1}}{_{B_2}}{_{\dots}}{_{B_n}}
&=&
\Omega^{-1}
\varphi{_{B_1}}{_{B_2}}{_{\dots}}{_{B_n}}.\label{167}
\ee
For  $\Omega=|\Omega|$ the massless fields are conformally invariant with conformal weight $-1$, which is the standard result. For a complex $\Omega$ the weight 
is given by (\ref{161}). 

\subsection{Penrose connection for $T_{abc}\neq 0$}
\label{Sec B}

If one insists on (\ref{167}) for a complex $\Omega$ one may follow the suggestion of Penrose \cite{P1983} and assume that $T_{abc}\neq 0$, with $\Pi_a=0$ in (\ref{157})--(\ref{158}),
\be
\Theta{_{a}}{_{B}}{^{X}}
&=&
\Upsilon{_{B}}{_{A'}}\ve{_{A}}{^{X}},\\
\Upsilon{_{a}}
&=&
\Omega^{-1}\partial{_{a}}\Omega,\\
\Phi{_{B_1}}{_{B_2}}{_{\dots}}{_{B_n}}
&=&
\Omega^{-1} \varphi{_{B_1}}{_{B_2}}{_{\dots}}{_{B_n}}.
\ee
Then
\begin{widetext}
\be
\Omega\nabla{_{a}}\Phi{_{B_1}}{_{B_2}}{_{\dots}}{_{B_n}}
&=&
\partial{_{a}}\varphi{_{B_1}}{_{B_2}}{_{\dots}}{_{B_n}}
\nonumber\\
&\pp=&
-
(\partial{_{a}}\ln\Omega)\varphi{_{B_1}}{_{B_2}}{_{\dots}}{_{B_n}}
-
(\partial{_{B_1}}{_{A'}}\ln\Omega)
\varphi{_{A}}{_{B_2}}{_{\dots}}{_{B_n}}
+
\dots
-
(\partial{_{B_n}}{_{A'}}\ln\Omega)
\varphi{_{B_1}}{_{B_2}}{_{\dots}}{_{A}}\label{170}
\ee
\end{widetext}
which is analogous to the right-hand side of (\ref{340}). 
Symmetry $\varphi{_{B_1}}{_{\dots}}{_{B_n}}=\varphi{_{(B_1}}{_{\dots}}{_{B_n)}}$ implies
\be
\Omega
\nabla{^{B_j}}{_{A'}}\Phi{_{B_1}}{_{\dots}}{_{B_j}}{_{\dots}}{_{B_n}}
&=&
\partial{^{B_j}}{_{A'}}
\varphi{_{B_1}}{_{\dots}}{_{B_j}}{_{\dots}}{_{B_n}}
\ee
so that the massless field is conformally invariant with weight $-1$, as in the real case $\Omega=|\Omega|$, but for the price of  non-vanishing torsion.

\subsection{Alternative connection for $T_{abc}\neq 0$}
\label{Sub C}

Once we decide on non-zero torsion, we may go back to
 (\ref{151})--(\ref{152}) and take the simple case of connections whose symmetric parts vanish,
\be
\Theta{_{a}}{_{(B}}{_{C)}}
=0
=
\bar\Theta{_{a}}{_{(B'}}{_{C')}}.
\ee
Then
\be
\Theta{_{a}}{_{B}}{_{C}}
&=&
\frac{1}{2}
\Om^{-1}\partial{_{a}}\Omega\,
\ve{_{B}}{_{C}},\label{174}
\\
\bar\Theta{_{a}}{_{B'}}{_{C'}}
&=&
\frac{1}{2}
\bar\Om^{-1}\partial{_{a}}\bar\Omega\,
\ve{_{B'}}{_{C'}},\label{175}
\ee
leads to covariant derivatives
\be
\nabla{_{a}}f{_{B}}
&=&
\big(
\partial{_{a}}
+
\partial{_{a}}\ln \Omega^{-\frac{1}{2}}
\big) f{_{B}},\\
\nabla{_{a}}f{_{B'}}
&=&
\big(
\partial{_{a}}
+
\partial{_{a}}\ln \bar\Omega^{-\frac{1}{2}}
\big) f{_{B'}},
\ee
with nontrivial torsion tensor
\be
T{_a}{_b}{_c}
&=&
-
\partial{_{a}}\ln|\Omega|
g{_{b}}{_{c}}
+
\partial{_{b}}\ln|\Omega|
g{_{a}}{_{c}}.
\ee
Infeld-van der Waerden connection satisfies $|\Omega|=1$, so its torsion vanishes and we are back to Sec.~\ref{Sec A} with $\Upsilon_a=0$.

The main advantage of our new form of connection can be seen in the formula linking $\nabla$ with $\partial$,
\begin{widetext}
\be
\Omega^{-\frac{n-k}{2}}\bar\Omega^{-\frac{m-l}{2}}\nabla{_{a}}
\big(\Omega^{\frac{n-k}{2}}\bar\Omega^{\frac{m-l}{2}}\varphi{_{B_1}}{_{\dots}}{_{B_n}}{_{B'_1}}{_{\dots}}{_{B'_m}}
{^{C_1}}{^{\dots}}{^{C_k}}{^{C'_1}}{^{\dots}}{^{C'_l}}\big)
&=&
\partial{_{a}}\varphi{_{B_1}}{_{\dots}}{_{B_n}}{_{B'_1}}{_{\dots}}{_{B'_m}}
{^{C_1}}{^{\dots}}{^{C_k}}{^{C'_1}}{^{\dots}}{^{C'_l}}.\label{main}
\ee
(\ref{main}) just links $\nabla{_{AA'}}$ with $\partial{_{AA'}}$ and does not involve transvection of $A$ with a field index.
For this reason,  (\ref{main}) holds true {\it independently of field equations\/} fulfilled by $\varphi$. 
This is why this type of covariant differentiation may be employed in the particular case of $m\neq 0$ fields.

Formulas
\be
\Phi{_{B_1}}{_{\dots}}{_{B_n}}{_{B'_1}}{_{\dots}}{_{B'_m}}
{^{C_1}}{^{\dots}}{^{C_k}}{^{C'_1}}{^{\dots}}{^{C'_l}}
&=&
\Omega^{\frac{n-k}{2}}\bar\Omega^{\frac{m-l}{2}}\varphi{_{B_1}}{_{\dots}}{_{B_n}}{_{B'_1}}{_{\dots}}{_{B'_m}}
{^{C_1}}{^{\dots}}{^{C_k}}{^{C'_1}}{^{\dots}}{^{C'_l}},\\
\Omega^{-\frac{n-k}{2}}\bar\Omega^{-\frac{m-l}{2}}\nabla{_{a}}
\Phi{_{B_1}}{_{\dots}}{_{B_n}}{_{B'_1}}{_{\dots}}{_{B'_m}}
{^{C_1}}{^{\dots}}{^{C_k}}{^{C'_1}}{^{\dots}}{^{C'_l}}
&=&
\partial{_{a}}\varphi{_{B_1}}{_{\dots}}{_{B_n}}{_{B'_1}}{_{\dots}}{_{B'_m}}
{^{C_1}}{^{\dots}}{^{C_k}}{^{C'_1}}{^{\dots}}{^{C'_l}}\label{main'},
\ee
when compared with the complicated expressions (\ref{340}) and (\ref{170}) show the degree of simplification and generality we obtain. Of particular interest is the case where $\Omega$ relates background Minkowski space with the universe defined by $\Psi_\tau(x)\neq 0$.

In the next section we will discuss the free Dirac equation with non-zero mass, but first let us cross-check some important special cases.
For $\ve_{AB}$ we have $n=2$, $m=k=l=0$, and we indeed get
\be
\Omega^{-1}\nabla{_{a}}
\big(\Omega\ve{_{B_1}}{_{B_2}}\big)
&=&
\partial{_{a}}\ve{_{B_1}}{_{B_2}}=0
\ee
because $\ve$ is independent of $x$. 
Analogously, for $\ve^{AB}$ we have $k=2$, $m=n=l=0$, 
\be
\Omega\nabla{_{a}}
\big(\Omega^{-1}\ve
{^{C_1}}{^{C_2}}\big)
&=&
\partial{_{a}}\ve
{^{C_1}}{^{C_2}}=0.
\ee
Of particular interest is the case of the world-vector field $x^a$ itself ($n=m=0$, $k=l=1$),
\be
\Omega^{\frac{1}{2}}\bar\Omega^{\frac{1}{2}}\nabla{_{a}}
\big(\Omega^{-\frac{1}{2}}\bar\Omega^{-\frac{1}{2}}
x{^{b}}\big)
&=&
x{^{b}}\partial{_{a}}\ln \Omega^{-\frac{1}{2}}\bar\Omega^{-\frac{1}{2}}
+
\partial{_{a}}x{^{b}}
+
x{^{b}}\partial{_{a}}\ln \Omega^{\frac{1}{2}}\bar\Omega^{\frac{1}{2}}
\\
&=&
\partial{_{a}}x{^{b}}=g{_a}{^b}.
\ee
Similar calculation yields
\be
\Omega^{-\frac{1}{2}}\bar\Omega^{-\frac{1}{2}}\nabla{_{a}}
\big(\Omega^{\frac{1}{2}}\bar\Omega^{\frac{1}{2}}
x{_{b}}\big)
&=&
g_{ab}.
\ee
The formulas are valid for any $\Omega$, complex or real. Actually, in the next Section we will see that the case $\Omega=|\Omega|$ is particularly interesting when it comes to massive fields.
\end{widetext}

\section{First-quantized Dirac equation}
\label{Sec int4}

Let us consider the first-quantized free Dirac's electron with mass $m$ as a test of the proposed description of conformal properties of massive fields.
For large $\tau$ what we expect is essentially the bispinor field $\big(\psi^A(x),\psi^{A'}(x)\big)$, $x\in\cal M$, which is scanned by means by the subspace of $\cal M$ defined by $\Psi_\tau(x)\neq 0$. This subspace looks `almost like a hyperplane' propagating toward the future. If we assume that a single Dirac electron does not influence the evolution of 
$\Psi_\tau(x)$, we can treat $\Psi_\tau(x)$ as a given solution of an empty universe Schr\"odinger equation that determines the flow of quantum time. 

Obviously, we do not discuss here the full dynamics with Hamiltonian (\ref{total H}) and second-quantized energy momentum tensor of the Dirac field (cf. sections 5.8-5.10 in \cite{PR}). Instead of discussing the influence of matter fields on the wave function of the universe, we try to understand why and how the concrete choice of $\Psi_\tau(x)$ may look like a conformal modification of geometry of space-time. 

2-spinor form of Dirac's equation for electron of mass $m$ is given in the background Minkowski space by \cite{PR} 
\be
\partial{_B}{_{A'}}\psi{^B} &=& -M \psi{_{A'}},\label{D1}\\
\partial{_A}{_{B'}}\psi{^{B'}} &=& -M \psi{_{A}},\label{D2}
\ee
where $M=mc/(\hbar\sqrt{2})$. (\ref{main}) implies 
\be
\Omega^{\frac{1}{2}}\nabla{_{a}}
\big(\Omega^{-\frac{1}{2}}\varphi
{^{C_1}}\big)
&=&
\partial{_{a}}\varphi{^{C_1}}, \quad \textrm{($n=m=l=0$, $k=1$),}\\
\bar\Omega^{\frac{1}{2}}\nabla{_{a}}
\big(\bar\Omega^{-\frac{1}{2}}\varphi
{^{C'_1}}\big)
&=&
\partial{_{a}}\varphi
{^{C'_1}},\quad \textrm{($n=m=k=0$, $l=1$),}
\ee
so Dirac's equation is transformed into
\be
\nabla{_B}{_{B'}}
\big(\Omega^{-\frac{1}{2}}\psi{^B}\big) &=& -M \Omega^{-\frac{1}{2}}\psi{_{B'}},
\label{D1'}\\
\nabla{_B}{_{B'}}
\big(\bar\Omega^{-\frac{1}{2}}\psi{^{B'}}\big) 
&=& 
-
M \bar\Omega^{-\frac{1}{2}}\psi{_{B}}\label{D2'}.
\ee
If $\Omega=|\Omega|$ then (\ref{D1'})--(\ref{D2'}) is just a conformally transformed (\ref{D1})--(\ref{D2}).

The link between conformal invariance and mass crucially depends on torsion of the connection. The result seems interesting in itself and deserves further studies.

The most natural choice of $\Omega$ then corresponds to (\ref{143}) if we additionally assume that the wave function of the universe $\Psi_\tau(x)$ is real and non-negative. Reality and non-negativity are preserved by (\ref{2''}). 

The conformally rescaled Dirac equation now reads
\be
\nabla{_B}{_{B'}}
\big(\sqrt{\Psi_\tau}\psi{^B}\big) &=& -M \sqrt{\Psi_\tau}\psi{_{B'}},
\label{D1''}\\
\nabla{_B}{_{B'}}
\big(\sqrt{\Psi_\tau}\psi{^{B'}}\big) 
&=& 
-
M \sqrt{\Psi_\tau}\psi{_{B}}\label{D2''},
\ee
with covariant derivatives
\be
\nabla{_{a}}f{_{B}}
&=&
\big(
\partial{_{a}}
+
\partial{_{a}}\ln \sqrt{\Psi_\tau}
\big) f{_{B}},\label{195}\\
\nabla{_{a}}f{_{B'}}
&=&
\big(
\partial{_{a}}
+
\partial{_{a}}\ln \sqrt{\Psi_\tau}
\big) f{_{B'}},\label{196}
\ee
and torsion
\be
T{_a}{_b}{_c}
&=&
\partial{_{a}}\ln \Psi_\tau
g{_{b}}{_{c}}
-
\partial{_{b}}\ln \Psi_\tau
g{_{a}}{_{c}}.
\ee
Covariant derivatives (\ref{195})--(\ref{196}) may be easily confused with standard modifications of $\partial_a$ by a local  $U(1)$ electromagnetic gauge transformation. The main difference is that the connection in (\ref{195})--(\ref{196}) is spin-dependent, i.e. depends on the spinor type of the field. So this is a true spin connection, unrelated to the notion of charge. Generalization of spinor connections to charged fields is described in Chapter 5 of  \cite{PR}. The same construction can be adapted here.

Let us note that the spinor indices have been raised and lowered by means of Minkowskian $\ve^{AB}$ and $\ve_{AB}$. 
This can be regarded as a logical inconsistency, which leads now to an alternative interpretation of (\ref{D1'})--(\ref{D2'}).

Indeed, we have introduced $\nabla_a$ by demanding (\ref{300})--(\ref{303}). Returning to (\ref{D1'})--(\ref{D2'}), but rewritten as
\be
\nabla{_B}{_{B'}}
\hat\psi{^B} &=& -M \Omega^{-\frac{1}{2}}\bar\Omega^{-\frac{1}{2}}
\underbrace{\bar\Omega^{-\frac{1}{2}}\psi{^{C'}}}_{\hat\psi{^{C'}}}
\underbrace{\bar\Omega\ve{_{C'}}{_{B'}}}_{\hat\ve{_{C'}}{_{B'}}},
\label{D1'''}\\
\nabla{_B}{_{B'}}
\hat\psi{^{B'}}
&=& 
-
M \bar\Omega^{-\frac{1}{2}}\Omega^{-\frac{1}{2}}
\underbrace{\Omega^{-\frac{1}{2}}\psi{^{C}}}_{\hat\psi{^{C}}}
\underbrace{\Omega\ve{_{C}}{_{B}}}_{\hat\ve{_{C}}{_{B}}}
\label{D2'''},
\ee
we obtain Dirac's equation with spinor indices lowered according to the rules of the universe, and not the ones of the background Minkowski space,
\be
\nabla{_B}{_{B'}}
\hat\psi{^B} &=& -\hat M \hat\psi{_{B'}},
\label{D1''''}\\
\nabla{_B}{_{B'}}
\hat\psi{^{B'}}
&=& 
-
\hat M \hat\psi{_{B}}
\label{D2''''}.
\ee
Here 
\be
\hat\psi{^B} &=& \Omega^{-\frac{1}{2}}\psi{^{B}},\\
\hat\psi{^{B'}} &=& \bar\Omega^{-\frac{1}{2}}\psi{^{B'}},\\
\hat\ve{_{C}}{_{B}} &=& \Omega\ve{_{C}}{_{B}},\\
\hat\ve{_{C'}}{_{B'}} &=& \bar\Omega\ve{_{C'}}{_{B'}},\\
\hat\psi{_B} &=& \hat\psi{^C}\hat\ve{_{C}}{_{B}}=\Omega^{\frac{1}{2}}\psi{_{B}},\\
\hat\psi{_{B'}} &=& \hat\psi{^{C'}}\hat\ve{_{C'}}{_{B'}}=\bar\Omega^{\frac{1}{2}}\psi{_{B'}},\\
\hat M &=& M \bar\Omega^{-\frac{1}{2}}\Omega^{-\frac{1}{2}}.
\ee
The last term is Higgs-like. Indeed, squaring the mass and employing (\ref{143}), we find
\be
\hat M^2 &=& M^2 \bar\Omega^{-1}\Omega^{-1}=M^2 |\Psi_\tau|^2.
\ee
Possible links between conformal rescalings and Higgs fields have been investigated in  \cite{FR1988,PawlowskiRaczka}, but typically with the implicit assumption of zero mass.  The present construction sheds new light on the problem and requires further studies.

$\Omega$ can in principle be complex, but 
$\Omega=|\Omega|=\Psi_\tau^{-1}$ is again the simplest choice:
\be
\hat\psi{^B} &=& \sqrt{\Psi_\tau}\psi{^{B}},\\
\hat\psi{^{B'}} &=& \sqrt{\Psi_\tau}\psi{^{B'}},\\
\hat\ve{_{C}}{_{B}} &=& \Psi_\tau^{-1}\ve{_{C}}{_{B}},\\
\hat\ve{_{C'}}{_{B'}} &=& \Psi_\tau^{-1}\ve{_{C'}}{_{B'}},\\
\hat\ve{^{C}}{^{B}} &=& \Psi_\tau\ve{^{C}}{^{B}},\\
\hat\ve{^{C'}}{^{B'}} &=& \Psi_\tau\ve{^{C'}}{^{B'}},\\
\hat\psi{_B} &=& \hat\psi{^C}\hat\ve{_{C}}{_{B}}=\psi{_{B}}/\sqrt{\Psi_\tau},\\
\hat\psi{_{B'}} &=& \hat\psi{^{C'}}\hat\ve{_{C'}}{_{B'}}=\psi{_{B'}}/\sqrt{\Psi_\tau},\\
\hat g_{ab} &=& |\Psi_\tau|^{-2}g_{ab},\\
\hat g^{ab} &=& |\Psi_\tau|^{2}g^{ab},\\
\hat M &=& M \Psi_\tau= M\sqrt{\mathscr{Z}}\chi_\tau.
\ee
$\mathscr{Z}=\max_x\{|\Psi_\tau(x)|^2\}$ is a renormalization constant. Effectively, the mass of the electron, as seen from the interior of the universe,  becomes renormalized and multiplied by a cutoff function.


\section{$1+1$ revisited}
\label{Sec 1+1}

This section summarizes all the essential steps of the construction on toy models in $(1+1)$-dimensional Minkowski space. Calculations are performed in hyperbolic coordinates but, as opposed to \cite{MCAP}, do not crucially depend on their properties. 

\subsection{Scalar product}

In hyperbolic coordinates,
\be
x^0= \mathtt{x}\cosh\xi,\quad
x^1 = \mathtt{x}\sinh\xi,
\ee
the scalar product reads
\be
\langle F|G\rangle
&=&
\int_{V_+} d^2x\, \overline{F(x^0,x^1)}G(x^0,x^1)\\
&=&
\int_{0}^\infty d\mathtt{x}\int_{-\infty}^\infty d\xi \, \mathtt{x}\overline{f(\mathtt{x},\xi)}g(\mathtt{x},\xi)
=
\langle f|g\rangle,\label{224}
\ee
where $f(\mathtt{x},\xi)=F(\mathtt{x}\cosh\xi,\mathtt{x}\sinh\xi)$, etc. 

\subsection{Dynamics of empty universe}

The dynamics is given by
\be
\Psi_{\tau}(x)
&=&
\Psi_{\tau_0}\left(
\sqrt{\mathtt x^2-(a_\tau)^2+(a_{\tau_0})^2}\cosh\xi,\sqrt{\mathtt x^2-(a_\tau)^2+(a_{\tau_0})^2}\sinh\xi
\right)\\
&=&
\psi_{\tau_0}\left(
\sqrt{\mathtt x^2-(a_\tau)^2+(a_{\tau_0})^2},\xi
\right)
=\psi_{\tau}(\mathtt{x},\xi), \quad\textrm{for $\mathtt x^2-(a_\tau)^2+(a_{\tau_0})^2>0$}
\ee
and
\be
\psi_{\tau}(\mathtt{x},\xi)=0, \quad\textrm{for $\mathtt x^2-(a_\tau)^2+(a_{\tau_0})^2\le 0$.}
\ee
Empty-universe Hamiltonian
\be
{\cal H}_0
&=&
-i
\frac{\ell^2}{2\mathtt x^2}
x^{\mu}
\partial _\mu
=
-i
\ell^2
\frac{\partial}{\partial(\mathtt{x}^2)}
\ee
implies that the dynamics acts by displacement in the $\mathtt{x}^2$ variable,
\be
\psi_{\tau}(\mathtt{x},\xi)
&=&
e^{-i\left((a_{\tau})^2-(a_{\tau_0})^2\right){\cal H}_0/\ell^2}
\psi_{\tau_0}(\mathtt{x},\xi)\\
&=&
e^{-\left((a_{\tau})^2-(a_{\tau_0})^2\right)\frac{\partial}{\partial(\mathtt{x}^2)}}
\psi_{\tau_0}(\sqrt{\mathtt{x}^2},\xi)
=
\psi_{\tau_0}\left(
\sqrt{\mathtt x^2-(a_\tau)^2+(a_{\tau_0})^2},\xi
\right),\quad 
\textrm{for $\mathtt x^2\geq (a_\tau)^2-(a_{\tau_0})^2$,}\label{234}
\ee
and 
\be
\psi_{\tau}(\mathtt{x},\xi)
&=& 0,\quad 
\textrm{for $\mathtt x^2< (a_\tau)^2-(a_{\tau_0})^2$.}
\ee
The parameter that plays a role of a `quantum time' is given by $(a_\tau)^2$. The simplest parametrization is 
$(a_\tau)^2=\ell^2\tau$.

\subsection{Group vs. semigroup}

The dynamics is unitary for any initial condition $\psi_{\tau_0}(\mathtt{x},\xi)$  if $\tau_0\to \tau$ is equivalent to translation by $(a_\tau)^2-(a_{\tau_0})^2$ {\it to the right\/} in the space of the variable $\mathtt{x}^2\in\mathbb{R}_+$. This is equivalent to $(a_\tau)^2-(a_{\tau_0})^2\geq 0$. 

\begin{figure}
\includegraphics[width=8 cm]{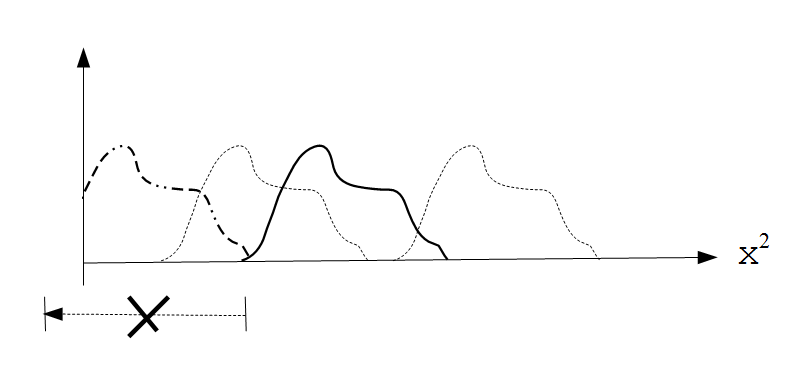}
\caption{All translations to the right and some translations to the left  are unitary, as opposed to those translations to the left that decrease the area under the curve. The unallowed translations cannot bring us to the negative axis of $\mathbb{R}$, yet they influence the norm of the state vector. Any shift to the right is unitary and reversible (local reversibility).  In this sense, global arrow of time coexists with local reversibility of time evolution. Note that the shifted variable is not $\mathtt{x}$ but 
$\mathtt{x}^2$ (for $n=1+1$), and $\mathtt{x}^n$ for arbitrary $n$. The shift in $\mathtt{x}^n$ shrinks the  timelike width of the membrane when plotted in plain $\mathtt{x}$-coordinates.}
\label{Fig2}
\end{figure}

Our dynamics is effectively given by a unitary representation of the semigroup of translations in 
$\mathbb{R}_+$. If the translation $\psi_{\tau_0}(\mathtt{x},\xi)\mapsto \psi_{\tau}(\mathtt{x},\xi)$ is to the right, the inverse translation to the left, $\psi_{\tau}(\mathtt{x},\xi)\mapsto \psi_{\tau_0}(\mathtt{x},\xi)$ is unitary as well (evolution is locally reversible). However, although all translations to the right are unitary, this is not true of all the translations to the left. 
The latter property automatically introduces a global arrow of time, in spite of local reversibility. Fig.~\ref{Fig2} illustrates these properties. 

\subsection{Unitarity of the semigroup}

For simplicity assume  $(a_{\tau_0})^2=\ell^2\tau_0=0$. One begins with
\be
\langle f_\tau|g_\tau\rangle
&=&
\int_{0}^\infty d\mathtt{x}\int_{-\infty}^\infty d\xi \, \mathtt{x}\overline{f_\tau(\mathtt{x},\xi)}g_\tau(\mathtt{x},\xi).
\label{239}
\ee
The dynamics is given by,
\be
f_\tau(\mathtt{x},\xi)
&=&
\left\{
\begin{array}{cl}
f_{0}\left(
\sqrt{\mathtt x^2-(a_\tau)^2},\xi
\right) & \textrm{for $(a_\tau)^2\le \mathtt x^2$}\\
0 & \textrm{for $0\le\mathtt x^2\le (a_\tau)^2$}
\end{array}
\right.
\label{240}
\ee
(and analogously for $g_\tau(\mathtt{x},\xi)$).
Inserting (\ref{240}) into (\ref{239}), and then changing variables $\mathtt{y}^2=\mathtt x^2-(a_\tau)^2$, we find
\be
\langle f_\tau|g_\tau\rangle
&=&
\frac{1}{2}
\int_{(a_\tau)^2}^\infty d(\mathtt{x}^2)\int_{-\infty}^\infty d\xi \, 
\overline{f_{0}\left(
\sqrt{\mathtt x^2-(a_\tau)^2},\xi
\right)}
g_{0}\left(
\sqrt{\mathtt x^2-(a_\tau)^2},\xi
\right)
\\
&=&
\frac{1}{2}
\int_{0}^\infty d(\mathtt{y}^2)\int_{-\infty}^\infty d\xi \, 
\overline{f_{0}\left(
\sqrt{\mathtt y^2},\xi
\right)}
g_{0}\left(
\sqrt{\mathtt y^2},\xi
\right)
=
\langle f_{0}|g_{0}\rangle.
\ee
It is clear that vanishing of (\ref{240}) for $0\le\mathtt x^2\le (a_\tau)^2$ is essential for the proof of 
$\langle f_{\tau}|g_{\tau}\rangle=\langle f_{\tau_0}|g_{\tau_0}\rangle$. Disappearance of the past becomes a  sine qua non condition for unitarity!

\subsection{Timelike width of the membrane}

Now assume an initial condition satisfying
\be
\psi_{0}(\mathtt{x},\xi)
&=& 0,\quad 
\textrm{for $\mathtt x \geq\Delta_{0}$,}\label{236}
\ee
for some $\Delta_{0}>0$. 
Accordingly, the initial wave function can be nonzero, $\psi_{0}(\mathtt{x},\xi)\neq 0$, only for $0<\mathtt x <\Delta_{0}$. Formula (\ref{234}) implies that 
 $\psi_{\tau}(\mathtt{x},\xi)\neq 0$, only for $0<\sqrt{\mathtt x^2-(a_\tau)^2} <\Delta_{0}$,
i.e.
\be
|a_\tau| <\mathtt x<\sqrt{(a_\tau)^2 +(\Delta_{0})^2}.
\ee
The timelike width of the membrane shrinks to zero with $a_\tau$ growing to infinity,
\be
\Delta_{\tau}
&=&
\sqrt{(a_\tau)^2 +(\Delta_{0})^2}-|a_\tau|
=
\frac{(\Delta_{0})^2}{\sqrt{(a_\tau)^2 +(\Delta_{0})^2}+|a_\tau|}
\underset{a_\tau\to\infty}{\longrightarrow}
0.
\ee
In $n$-dimensional Minkowski space the effect is even more pronounced as the shifted variable is $\mathtt{x}^n$.

\subsection{Spectral properties of the Hamiltonian}

The eigenvalue problem is
\be
-i
\ell^2
\frac{\partial}{\partial(\mathtt{x}^2)}
f_E(\mathtt{x}) &=& E f_E(\mathtt{x}),\\
f_E(\mathtt{x})
&=&
f_E(0)e^{iE \mathtt{x}^2/\ell^2}.\label{233}
\ee
Scalar product (\ref{224}) implicitly involves integration $\int_0^\infty d(\mathtt{x}^2)$, over the same variable that occurs in (\ref{233}). Spectral theorem reduces here to Fourier analysis of wave-packets whose supports are subsets of $\mathbb{R}_+$. Fourier-transform artefacts at 0, such as the Gibbs phenomenon, do not occur because we consider wave packets continuous (and vanishing) at $\mathtt{x}=0$. Eigenvectors of ${\cal H}_0$ (plane waves) are complete.  Eigenvalues $E$ are given by arbitrary real numbers (the Hamiltonian is unbounded from below). The same discussion applies to Minkowski spaces of any dimension $n$.

\subsection{Interaction with matter: Shape dynamics as an example}

Let us  consider some toy model of a universe filled with matter. For illustrative purposes, the matter content can be described by a discrete degree of freedom $A$. The wave function is $\psi_\tau(\mathtt{x},\xi,A)$, with total Hamiltonian
\be
i\dot \psi_\tau(\mathtt{x},\xi,A)
&=&
{\cal H}\psi_\tau(\mathtt{x},\xi,A)
\\
&=&
-i
\ell^2\frac{\partial}{\partial(\mathtt{x}^2)}\psi_\tau(\mathtt{x},\xi,A)
+
\sum_B{\cal H}_1(\mathtt{x},\xi){_{A}}{_B}\psi_\tau(\mathtt{x},\xi,B).
\ee
Following Barbour's shape dynamics \cite{Barbour} let us assume that interaction depends solely on the shape variable $\xi$,
\be
i\dot \psi_\tau(\mathtt{x},\xi,A)
&=&
-i
\ell^2\frac{\partial}{\partial(\mathtt{x}^2)}\psi_\tau(\mathtt{x},\xi,A)
+
\sum_B{\cal H}_1(\xi){_{A}}{_B}\psi_\tau(\mathtt{x},\xi,B).
\ee
Our shape dynamics involves interaction Hamiltonian analogous to the one from (\ref{total H}),
\be
{\cal H}_1=I\otimes \int_{-\infty}^\infty d\xi |\xi\rangle\langle\xi|\otimes {\cal H}_1(\xi)
\ee
Separating variables, $\psi_\tau(\mathtt{x},\xi,A)= f_\tau(\mathtt{x})g_\tau(\xi,A)$, we obtain
\be
i\dot f_\tau(\mathtt{x})
&=&
-i
\ell^2\frac{\partial}{\partial(\mathtt{x}^2)}f_\tau(\mathtt{x}),\\
i\dot g_\tau(\xi,A)
&=&
\sum_B{\cal H}_1(\xi){_{A}}{_B}g_\tau(\xi,B).
\ee
The solution
\be
\psi_\tau(\mathtt{x},\xi,A)
&=&
\left\{
\begin{array}{cl}
f_0\left(
\sqrt{\mathtt x^2-\ell^2\tau}\right)e^{-i\tau{\cal H}_1} g_0(\xi,A) & \textrm{for $\ell^2\tau\le \mathtt x^2$}\\
0 & \textrm{for $0\le\mathtt x^2\le \ell^2\tau$}
\end{array}
\right.
\label{246}
\ee
represents an entangled shape-matter state,
\be
|\psi_\tau\rangle &=& \int_0^\infty d\mathtt{x}\,\mathtt{x} 
f_\tau(\mathtt{x})
|\mathtt{x}\rangle\otimes
\sum_A\int_{-\infty}^\infty d\xi\, 
g_\tau(\xi,A)
|\xi\rangle\otimes |A\rangle,
\ee
while matter alone is described by the reduced density matrix
\be
\rho^M_\tau
&=&
\sum_{AB}\int_{-\infty}^\infty d\xi\, 
g_\tau(\xi,A)\overline{g_\tau(\xi,B)}|A\rangle\langle B|.
\ee
Wave function of the universe is influenced by the presence of matter. Probability density of the universe alone is given by
$\sum_A|\psi_\tau(\mathtt{x},\xi,A)|^2$,
and depends on the form of ${\cal H}_1(\xi)$. With initial condition analogous to (\ref{236}),
\be
f_0(\mathtt{x})
&=& 0,\quad 
\textrm{for $\mathtt x \geq\Delta_{0}$,}\label{236'}
\ee
we obtain probability density that vanishes for $\mathtt{x}\not\in\big[\ell\sqrt{\tau},\sqrt{\ell^2\tau+\Delta_0^2}\big]$.

It should be emphasized that the evolution parameter $\tau$ is huge --- it counts out cosmic time since $\tau=0$. The parameter we are dealing with in physical applications  corresponds to an infinitesimal increase $\tau\mapsto \tau+\Delta \tau$ (even if `infinitesimal' means in this context a million years). It is therefore justified to write
\be
\rho^M_{\tau+\Delta \tau}
&\approx&
\rho^M_{\tau}
-i\Delta \tau 
\sum_{ABC}\int_{-\infty}^\infty d\xi\, 
\left(
{\cal H}_1(\xi){_{A}}{_C}g_\tau(\xi,C)\overline{g_\tau(\xi,B)}
-
g_\tau(\xi,A)\overline{g_\tau(\xi,C)}{\cal H}_1(\xi){_{C}}{_B}
\right)|A\rangle\langle B|.
\ee
Assuming that influence of a small material system (a molecule, say) on the wave function of the universe is negligible, we can unentangle $\xi$ and $A$, $g_\tau(\xi,A)\approx g_\tau(\xi)\phi_\tau(A)$, so that
$\rho^M_\tau=|\phi_\tau\rangle\langle \phi_\tau|$, and
\be
\rho^M_{\tau+\Delta \tau}
&\approx&
\rho^M_{\tau}
-i\Delta \tau 
\sum_{ABC}\int_{-\infty}^\infty d\xi\, |g_\tau(\xi)|^2
\left(
{\cal H}_1(\xi){_{A}}{_C}\phi_\tau(C)\overline{\phi_\tau(B)}
-
\phi_\tau(A)\overline{\phi_\tau(C)}{\cal H}_1(\xi){_{C}}{_B}
\right)|A\rangle\langle B|\\
&=&
\rho^M_{\tau}
-i\Delta \tau 
[{\cal H}_\tau^M,\rho^M_{\tau}],
\ee
where we have introduced the effective matter Hamiltonian
\be
{\cal H}_\tau^M
&=&
\int_{-\infty}^\infty d\xi\, |g_\tau(\xi)|^2{\cal H}_1(\xi)
=
\int_{V_+}d^2x\,|\Psi_\tau(x)|^2{\cal H}_1(x).
\ee
This is precisely the Hamiltonian (\ref{124'}) we have arrived at by heuristic considerations. The evolution equation that represents evolution of small amounts of matter thus takes the usual von Neumann form
\be
i\dot \rho^M_{\tau} &=& [{\cal H}_\tau^M,\rho^M_{\tau}].
\ee
${\cal H}_\tau^M$ is $\tau$-dependent (integration is over a time dependent  domain), but at time scales available in quantum mechanical experiments it can be treated as time independent, ${\cal H}_{\tau+\Delta\tau}^M\approx{\cal H}_\tau^M$.

What we regard as a total Hamiltonian in our standard quantum mechanics or field theory turns out to be an interaction part of a true total Hamiltonian that includes the universe itself. From the point of view of matter alone, the wave function of the universe appears in a role of a `cutoff function', regularizing integrals over matter fields. The fact that timelike thickness $\Delta_\tau$ associated with $|\Psi_\tau(x)|^2$ shrinks to 0 is responsible for effectively 3D forms of 4D integrals 
occurring in (\ref{124'}) for late $\tau$s.


\section{Assumptions in a nutshell}

Similarly to Cort\'azar's  {\it Hopscotch\/}, our article can be read according to two different sequences of sections. The present one could become Section~II, while the previous one could play a role of Section~III (or the other way around). We will first concentrate on physical  intuitions behind our construction, and then sketch possibilities of some generalizations beyond the simple Minkowskian framework.

\subsection{Physical assumptions}

We treat $(1+3)$-dimensional space-time in exact analogy to 3-dimensional configuration space in nonrelativistic quantum mechanics. A point in a universe can exist in superposition of different locations, described by a $\tau$-dependent wave function $\Psi_\tau(x)$. A true universe may be regarded as a collection of $N$ such points, in analogy to $N$-particle systems in nonrelativistic quantum mechanics. What we do in the paper is essentially a 1-particle description, but an extension to 
$\Psi_\tau(x_1,\dots,x_N)$, $x_j\in V_+\subset\mathbb{R}^4$, based on (\ref{132_}), is worthy of further studies. A similar formalism but in momentum space was discussed in \cite{MC1,MC2,MC3,MC4,MC5}, with the conclusion that two types of $N\to\infty$ limits are physically meaningful. One plays a role of a weak law of large numbers, the other is interpreted as a thermodynamic limit. Such a perspective is conceptually close to the idea of a causal sets of discrete points in space-time \cite{Sorkin}, with space-times as their continuum limits, but in a  version involving wave packets instead of points (instead of a classical point we have a wave packet that represents a pointlike object, like in a matter-wave interferometer). 

The coupling (\ref{total H}) between space-time and matter is analogous to Hamiltonians occurring in the formalism of quantum time as proposed by Page and Wooters \cite{PW,GLM}. Again, a momentum-space analogue of such a `quantum time' structure can be found in \cite{MC1,MC2,MC3,MC4,MC5}.

The universe wave packet $\Psi_\tau(x)$ is extended in timelike directions by a nonzero width $\Delta_\tau$. A similar case occurs for the Chern-Simons time \cite{Smolin1994,Smolin2019}, although technically the Chern-Simons formalism is completely unrelated to what we propose. 

Popular explanations of general relativistic expansion of the Universe often employ a metaphor of an inflating balloon, meant to represent an expanding 3-dimensional submanifold of 4-dimensional space-time. The main intuition behind our formalism is similar, only the purely mathematical 3-dimensional submanifold is replaced by a finite-thickness membrane which resembles a true balloon. As the balloon expands its density decreases --- however, by density we mean the density of probability. The fabric of our universe is completely quantum.

In systematization proposed by Rovelli \cite{Rovelli} what we discuss is neither a global presentism, nor a static eternalism. At late $\tau$s, that is when $\Delta_\tau\approx 0$, we can speak of an approximately global approximate presentism.

Another important guiding principle behind our formalism is the correspondence principle with the usual quantum mechanics and field theory. At `late times' (of the order of 13-14 billion years) our new theory should reduce to something more standard, at least within sufficiently small neighborhoods of our labs (here `small' means `of the Galaxy size', say). We expect that matter Hamiltonians should be approximately time independent at least at time scales $\Delta t$ negligible with respect to 13-14 billion years. Only the full Hamiltonian is exactly independent of $\tau$. 
At corresponding size scales, volume of integration of matter-field Hamiltonian densities should be approximately flat, due to  negligible corrections to $d^3x$ arising from the curvature of proper-time hyperboloid at very late times.

We base the whole analysis on flat Minkowskian backgrounds, but it seems an analogous discussion could be performed in space-times that are only conformally Minkowskian, simply by augmenting formulas (\ref{136,})--(\ref{138,}) and (\ref{143})--(\ref{146}) by additional conformal factors. Alternatively, a conformally Minkowskian space-time should be first conformally transformed into the Minkowski space, then the construction would follow the lines we have discussed in the paper, and finally the result should be conformally transformed back. The formalism that seems especially suitable from our perspective is Barbour's shape dynamics \cite{Barbour} due to its natural separation of `time' and `space' variables discussed at the end of the  preceding section.

Coupling between matter and `geometry' is described by total Schr\"odinger-picture Hamiltonian ${\cal H}={\cal H}_0+
{\cal H}_1$ that involves a free part ${\cal H}_0$ responsible for expansion of the empty universe. The more standard form of matter-geometry interaction occurs only at an approximate level, if we treat $\Psi_\tau(x)$ as a background field which is not influenced by matter. So all the considerations involving formulas such as $|\Psi_\tau(x)|^2=\sqrt{|g_\tau(x)|}$ or the like, should be regarded as semiclassical.

\subsection{Mathematical assumptions}

First of all, a universe is represented by a subset of space-time defined by $\Psi_\tau(x)\neq 0$. The world-vector $x^a$ belongs to the future cone of some fiducial world-vector $X^a=0$, in some $n$-dimensional Minkowski space with signature $(+,-,\dots,-)$.
In principle, one could replace the fiducial world-vector $X^a$ by a point $X$ in some manifold. The field $\Psi_\tau(x)$ would then belong to a fiber over $X$. The boundary condition is: $\Psi_\tau(x)=0$ if $x^a$ is not future-timelike. In all the examples we assume that the support of 
$\Psi_\tau(x)$ is compact, which can be weakened if needed. We assume that $\Psi_\tau(x)$ is a complex scalar field, square-integrable with respect to $d^nx$, but any spinor field would do as well. The dynamics is given by a semigroup of translations in $\mathtt{x}^n$, where $n$ is the dimension of the background Minkowski space and $\mathtt{x}^2=g_{ab}x^ax^b$. 

This is one of the points that can be easily generalized beyond Minkowskian backgrounds. Indeed, it suffices to replace the Minkowskian $g_{ab}$ by a more general $g_{ab}$, provided a global foliation parametrized by $\mathtt{x}$ exists. The shape dynamics is a natural candidate for such generalizations. The dynamical semigroup would still be the one of translations of the $n$th power of $\mathtt{x}$. Links between shape dynamics and the new framework are intriguing and worthy of a detailed study.

\acknowledgments

Calculations were carried out at the Academic Computer Center in Gda{\'n}sk. The work was supported by the CI TASK grant `Non-Newtonian calculus with interdisciplinary applications'.

\end{document}